# The Emergence of Photonic Crystalline Order and Time-Series Dynamics in NaCl Droplet Deposition

Grzegorz S. Żmija[1], Grzegorz Cios[2], Benedykt R. Jany[1*]

[1]Marian Smoluchowski Institute of Physics, Faculty of Physics, Astronomy and Applied Computer Science, Jagiellonian University, ul. prof. Stanisława Lojasiewicza 11, 30-348 Krakow, Poland
[2]Academic Centre for Materials and Nanotechnology, AGH University of Krakow, al. A Mickiewicza 30, 30-059 Krakow, Poland

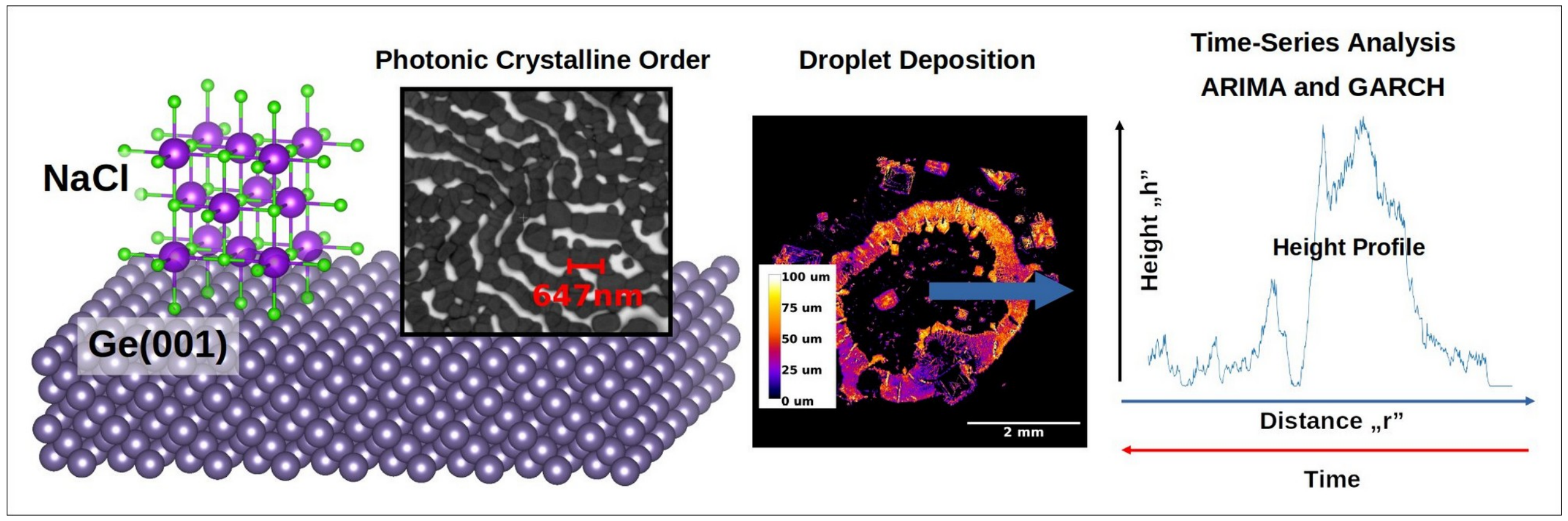

Keywords: Droplet Deposition, Photonic Crystal, Time-Series Analysis, NaCl, Germanium

## Abstract

Crystallization during droplet evaporation gives rise to complex, self-organized structures, yet the mechanisms underlying the emergence of ordered functional phases remain poorly understood. In this study, we present a comprehensive, multi-scale investigation into the crystallization dynamics of NaCl during droplet evaporation on a germanium (001) substrate, relevant for its IR applications. Through systematic microscopic characterization, we identify the formation of diverse microstructures, including 1D photonic crystal nanostructures formed within hybrid crystal-glass photonic system. To enable quantitative comparison across experimental conditions, we introduce the NaCl equivalent height as a unified metric to describe and classify the evolution of crystalline morphology. Our results reveal that diffusion anisotropy, rather than growth kinetics, primarily governs the maximal attainable structure size. Quantitative thin film interference analysis demonstrates the presence of discrete thickness layers in the film. Controlled evaporation experiments yield homogeneous crystallization patterns across the entire droplet area, facilitating the emergence of ordered photonic structures. Time-series dynamics analysis of height profiles uncovered the spatiotemporal evolution of the crystallization front, providing insights into the details of underlying physical mechanisms. Together, these results establish a robust experimental framework for understanding and predicting crystallization behavior in evaporating droplets, with potential applications in materials synthesis, photonics, and microscale pattern formation.

* corresponding author e-mail: benedykt.jany@uj.edu.pl

## Introduction

Sodium chloride (NaCl) was one of the first chemical compounds observed and described by humans. It is highly soluble in water (360 $dm^3$ per kg under standard conditions), non-toxic, hygroscopic, neutral in pH, white in color, and translucent in both visible and infrared light[1]. NaCl exists as a crystalline solid with a face-centered cubic lattice (space group $Fm\bar{3}m$), exhibiting a density of 2.17g/$cm^3$. The compound has a high melting point, with a triple point temperature of T=1073.8K, and is difficult to melt under normal conditions[2]. Industrially, NaCl plays a pivotal role as a dietary additive and as a chemical reagent used, for example, in the Mannheim process[3], the Solvay process[4], and in the production of polyvinyl chloride (PVC) polymers[5]. NaCl has also many different unique applications and interesting properties so it is commonly used in different fields of science and industry[6,7,8,9,10,11]. Germanium (Ge), alongside silicon (Si) and gallium arsenide (GaAs), is one of the most important and widely used semiconductor materials[12]. The element was discovered and characterized by Clemens Winkler in 1886, although Dmitri Mendeleev had already predicted its properties as early as 1869. Germanium is a crystalline solid with a diamond cubic structure (space group $Fd\bar{3}m$), displaying a silvery-gray appearance and a density of 5.33 g/$cm^3$. Under ambient conditions, it forms a thin native oxide layer (GeOx). Due to its narrow energy bandgap, approximately 0.67eV at room temperature, germanium is extensively employed in radiation detectors and photovoltaic devices. It also exhibits favorable optical properties in the visible and infrared regions[13]. The lattice constant of germanium under standard conditions is 5.658Å[14], while that of NaCl is 5.640Å[15]. This minimal difference in lattice parameters enables epitaxial growth, wherein a new crystalline layer grows with a well-defined orientation on a substrate. This phenomenon is exploited in various thin-film deposition techniques, particularly molecular beam epitaxy (MBE). Despite its advantages, such as high structural quality of the resulting films, MBE is limited by a low deposition rate (a few nanometers per minute) over limited areas and the requirement of ultra-high vacuum (UHV) environments. On a germanium substrate, sodium chloride can grow in multiple orientations, depending on the surface termination and deposition temperature. For instance, NaCl(100) has been observed on both Ge(100) and Ge(111) surfaces[16]. Thin NaCl layers have recently attracted considerable interest in industry as sacrificial substrates for the large-scale growth and processing of various functional materials, including gallium arsenide (GaAs)[17] and metallic nanostructures[18,19]. Gallium arsenide possesses direct bandgap, making it highly valuable in scientific research and technological applications, particularly in high-efficiency solar cells and radiation detection. The prevailing method for GaAs fabrication, MBE on III-V semiconductor layers, faces two major challenges. First, deposition substrates are typically not reusable, leading to manufacturing costs that can reach up to 30% of the final device's value[17]. Second, clean separation of the deposited GaAs layer from the substrate is extremely difficult and often results in significant degradation of both the film and the underlying substrate. These limitations can be overcome by first depositing a thin epitaxial NaCl layer, followed by the growth of GaAs on top. After completion, NaCl can be dissolved in water, enabling clean and non-damaging removal of the sacrificial layer[17]. A similar approach is applicable to other deposition techniques, such as physical vapor deposition (PVD) of metals. Furthermore, the high melting point of NaCl (1073.8 K) allows the substrate to be heated to temperatures exceeding 700°C, far beyond the thermal stability of polymer-based substrates, enabling the formation of complex nanostructures. Monocrystalline NaCl is widely used in optical systems, especially as a vacuum window for infrared radiation, owing to its excellent transmission properties in this spectral range. However, its hygroscopic nature remains a primary limitation, leading to degradation through a cyclic process involving dissolution, surface diffusion, and recrystallization of thin layers. Controlling these processes offers potential pathways to extend device lifetimes, enable self-regenerative systems, and develop novel manufacturing techniques for optical components[20]. Epitaxial NaCl layers also serve as sacrificial layers in the production of plastic and organic electronic devices[21]. In such applications, expensive and difficult-to-manufacture substrates, such as c-sapphire, are coated with a thin NaCl layer that replicates the desired surface structure. After the deposition of functional layers, the NaCl layer is selectively dissolved in water, allowing the final device to be easily released without damaging the underlying substrate, thereby significantly reducing production costs. Additionally, due to its very low solubility in nonpolar solvents, NaCl does not interfere with standard chemical processing methods for organic compounds, making it compatible with a wide range of synthesis procedures. Understanding the fundamental mechanisms governing the formation and stability of thin NaCl layers is essential for its effective use as a working agent or energy storage medium in high-temperature systems. However, NaCl's pro-corrosive behavior toward common metal alloys presents a significant challenge that must be carefully managed in practical applications[22].

Here we performed a comprehensive, multi-scale investigation into the crystallization dynamics of sodium chloride (NaCl) during droplet evaporation on a germanium (Ge) substrate. By combining advanced materials science, microscopy characterization together with Quantitative Thin Film Interference Analysis, and sophisticated time-series analysis, we uncovered both the microscopic mechanisms and macroscopic behaviors that govern the formation of complex, functional crystalline patterns, including coffee-ring-like structures and hybrid photonic phases.

## Materials and Methods

*NaCl Droplet Deposition*

The deposition experiments were performed on Germanium (001) epi-ready single-site polished wafer substrates, N-type undoped (MTI Crystals). The wafers were cut into square-like pieces with sample edges oriented along the <110> direction and cleaned with isopropanol in ultrasonic bath. A 99.5% NaCl AR/ACS (Loba Chemie, CAS: 7647-14-5) was used. The NaCl saturated solution was prepared by dissolving NaCl in distilled water until no further salt could dissolve. The appropriate NaCl solutions were derived from the volume fraction of the saturated solution and distilled water to ensure high repeatability and high accuracy[23]. For this purpose, a precise 10–100μL micro-pipette (Chem Land) was used, certified according to ISO 8655 / DIN 12650. The certificate specifies the following parameters: for a 25μL volume, the average delivery was 24.95μL, with an inaccuracy of −0.18% and an imprecision of 0.29%. Before droplet deposition, the germanium substrates were cleaned to remove surface oxides by rinsing for 1 minute in a 1 mol/L citric acid solution prepared using 99% citric acid (Sigma Aldrich). The substrates were subsequently rinsed with distilled water and isopropanol. This cleaning procedure ensures an oxide-free germanium surface, as demonstrated by G.Collins et al.[24]. NaCl droplet solution deposition on the clean germanium (001) sample substrates was carried out using a precise micro-pipette with a volume range of 0.1–2.5μL (Chem Land), certified according to ISO 8655 / DIN 12650. The certificate provides the following parameters: for a 1.00μL volume, the average delivery was 1.00μL, with an inaccuracy of 0.45% and an imprecision of 0.50%. For all experiments, unless otherwise stated, a droplet volume of 2.5μL was used. After droplet deposition, the samples were dried under constant ambient conditions in a petri dish on a paper disk in fume hood (average air ventilation of around ~700$m^3$/h). In total around 40 samples were prepared.

*Sample Characterization*

The samples were characterized by Light Microscopy using a digital microscope Delta Optical Smart 5MP PRO setup, as described in B.R. Jany[25], and High Magnification (~950x) Bright Field and Dark Field Microscopy (Dino Lite Edge AM7515MT8A). A total of ~200 microscopy images were collected and subsequently analyzed using 2D Autocorrelation Function (ACF) analysis with the free software Gwyddion[26]. To correct the data for inhomogeneous lighting conditions prior to ACF analysis, a third-degree polynomial background was subtracted.

Quantitative Thin Film Interference analysis, implemented in Python, was used to determine the thickness of the formed thin film by quantitative comparison of the Bright Field interference colors with a reference model of NaCl on Germanium (details provided in the manuscript text).

Time-series analysis using ARIMA and GARCH models was performed on height profiles using the statsmodels[27] and arch[28] Python libraries.

XRD measurements were performed using a Panalytical Empyrean diffractometer in the standard Bragg-Brentano theta-2theta geometry, employing a copper anode with a voltage of 40 kV and a current of 40 mA. A nickel filter was used during the measurements to suppress the $CuK_{\beta}$ contribution.

SEM EBSD measurements were performed using a Thermo Fisher Scientific Helios 5 pFIB microscope equipped with an Oxford Instruments Symmetry S3 EBSD detector operating in “Speed 1” mode with a diffraction pattern resolution of 156 × 128 pixels. The microscope was operated at an accelerating voltage of 10 kV and a beam current of approximately 13 nA. EBSD datasets were acquired using Aztec software version 6.2. Diffraction images were stored for subsequent refinement against dynamically simulated diffraction patterns using the MapSweeper module implemented in Aztec Crystal version 4.0.

## Results

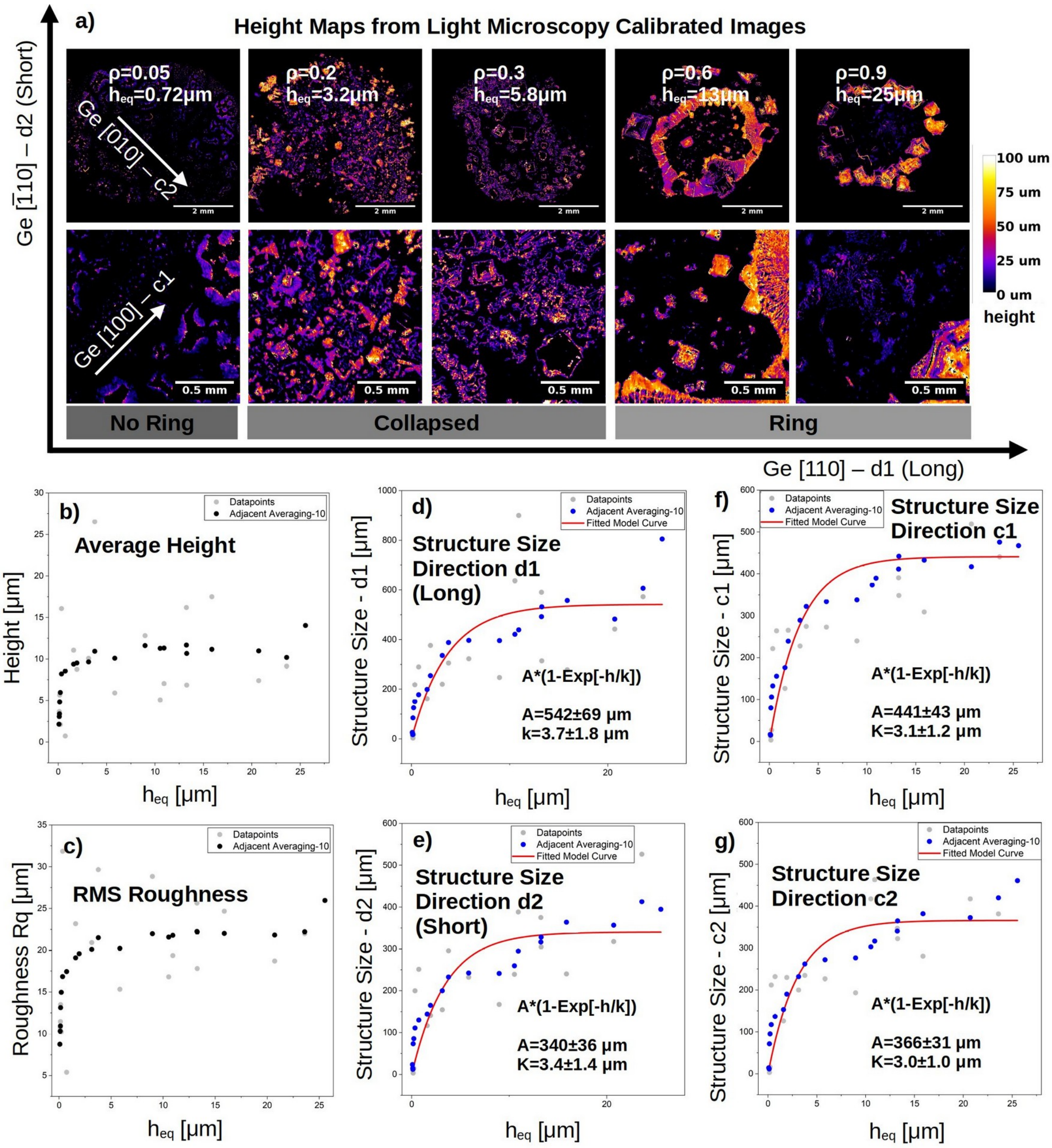

*Figure 1: Droplet deposition of NaCl on a Germanium (001) surface. a) Height maps derived from light microscopy calibrated images for different deposition conditions (ρ - concentration as a fraction of the saturated solution, $h_{eq}$ - NaCl equivalent height). Three distinct regions are visible: coffee ring region, collapsed coffee ring, and no coffee ring. Crystallographic directions d1 (Long), d2 (Short), c1, and c2 are indicated. b) and c) Average height and corresponding Root Mean Square (RMS) roughness of the formed structures. It is observed that a decrease in $h_{eq}$ leads to a corresponding decrease in both height and roughness. d) through g) Structure sizes (from autocorrelation analysis) as a function of $h_{eq}$ along directions d1, d2, c1, and c2, together with the growth model fit and parameters (A - residual structure size, k - growth length). It is observed that structures in the d1 (Long) direction exhibit the highest residual size (A), while the residual sizes for c1 and c2 are similar in the range of the extracted errors. The growth length factors (k) are similar across all directions. This indicates the highest diffusion strength along d1 direction.*

In Figure 1, the results of Light Microscopy (LM) imaging of droplet deposition of NaCl on a Germanium (001) surface are presented. The germanium substrate samples are cut such that the germanium <110> directions are parallel to the sample edges. Figure 1a) shows height maps derived from light microscopy height-calibrated images. Sample cross-sectional measurements were used to determine the height of the structures, and these were compared with the intensity of the CIE Lab Lightness channel in the LM images to establish a height versus intensity calibration. This calibration was fitted using the inverse Lambert-Beer law, as our samples consist of transparent salt that masks the dark germanium substrate. Finally, effective data based parametrization which translate Lightness into height was derived. For further details, see Figure S2 in the Supporting Information. This calibration was subsequently applied to convert LM images into height maps.

Different deposition conditions are indicated: ρ - the concentration of NaCl as a fraction of the saturated NaCl solution, and $h_{eq}$ - the NaCl equivalent height, defined as the thickness of NaCl material that would uniformly cover the entire droplet surface area. The equivalent height is calculated using the equation: $h_{eq}=\frac{V \rho g_r l}{S g_s}$ ,where V is the droplet volume, $g_r$ is the density of the NaCl-saturated solution, l is the solubility of NaCl, S is the droplet surface area, and $g_s$ is the density of NaCl. The $h_{eq}$ serves the same purpose as material coverage in standard MBE deposition and accounts for variations in experimental conditions such as surface area and solution concentration, acting as an ordering parameter. Three distinct regions are visible: a coffee ring formation region for high $h_{eq}$ (higher than ~10μm), an intermediate transitional region characterized by collapsed coffee rings (~0.9-10μm), and a region with no coffee ring formation (less then ~0.9μm). The performed XRD measurements clearly confirms, as expected[16], the epitaxial relation NaCl(001)//Ge(001) for all the regions, see Figure S3 in Supporting Information.  It is seen that when $h_{eq}$ decreases, both the average height and the average RMS roughness also decrease, see Fig. 1b)–c). We also noticed that the germanium surface wetting by the deposited droplet (contact angle - CA) changes with NaCl concentration being low (high CA) for high NaCl concentration and high (low CA) for low concentrations.

On a clear germanium surface at room temperature, a spontaneous surface reconstruction in the form of atomic wires along the <110> direction with a double atomic spacing appears[29]. These atomic wires represent the paths of the easiest and fastest material diffusion. Energetically, two possible scenarios exist: scenario A, where atomic terrace edges are formed parallel to the Ge[110] direction, and scenario B, where atomic terrace edges are formed parallel to the Ge[100] direction, for details, see Figure S1 in the Supporting Information. Crystallographic directions d1 Ge[110] (long), d2 Ge[−110] (short), c1 Ge[100], and c2 Ge[010] are indicated. The 2D Autocorrelation

Function (ACF) analysis on LM images was performed to determine the structure sizes in different crystallographic directions by extracting line profiles from the 2D ACF. The structure size is evaluated as the full width at half maximum (FWHM) of the central ACF peak. To experimentally distinguish between d1 Ge[110] and d2 Ge[−110], as well as between c1 Ge[100] and c2 Ge[010], the maximum size along the two Ge <110> directions is assigned as d1 (long) and the minimum size along Ge <110> directions as d2 (short). Similarly, the maximum size along the Ge <100> directions is assigned as c1, and the minimum size among the Ge <100> directions is assigned as c2. Figure 1d)-g) shows the extracted structure sizes as a function of $h_{eq}$ along the directions d1, d2, c1, and c2. It is seen that, for all cases, as $h_{eq}$ increases, the structure sizes "M" increase rapidly at first and then saturate at a fixed size value. To rationalize this behavior, we assume a first-order saturation growth model $\frac{dM}{dh_{eq}}=\frac{1}{k}(A-M)$ which yields $M(h_{eq})=A*[1-\exp(-h_{eq}/k)]$ . Here, A is the asymptotic maximum structure size and k a characteristic growth length. Since $h_{eq}$ is inversely proportional to the droplet surface area S, the growth is effectively limited by surface availability, consistent with a surface-limited growth mechanism. The model was fitted to the data. It is observed that structures in the d1 (Long) direction exhibit the highest residual size (A=549±69μm), while the residual sizes for c1 (A=441±43μm) and c2 (366±31μm) are similar within the range of the extracted errors. The larger residual size along the d1 direction, which corresponds to the dimer row direction on the Ge(001) surface, is consistent with enhanced surface diffusion along this direction. In contrast, the similar values of the characteristic growth length k ~3-4μm for all directions indicate that the underlying growth dynamics and rate limiting mechanisms are identical, with diffusion anisotropy primarily affecting the maximal attainable structure size rather than the growth kinetics. Furthermore, the structure size serves as a reliable indicator of the strength and directionality of surface diffusion.

Next we look into the details of the formed structures by high magnification Bright Field (BF) and Dark Field (DF) reflected light microscopy, as shown in Figure 2. In addition to the large NaCl crystals previously visible in LM, we now observe a finer structure of NaCl formed as thin films. We were able to distinguish six different structure morphologies, labeled A to F, as depicted in Fig. 2a). Similarly, we performed 2D ACF size analysis on the BF images. The structure size ratios along the directions d1/d2 versus c/d2 (where c is the average of c1 and c2 sizes) are shown in Fig. 2b), together with the corresponding parallel coordinates plot (c–d2–d1) in Fig. 2c). It is seen that different regions on the plots, corresponding to different strengths of material diffusion during structure growth, match with the distinct structure types A–F, indicated by different colors. These regions clearly show variations in diffusion behavior, ranging from weak to strong, and each color-coded zone consistently associates with a specific morphological signature, suggesting a direct link between diffusion anisotropy and the resulting structural form.

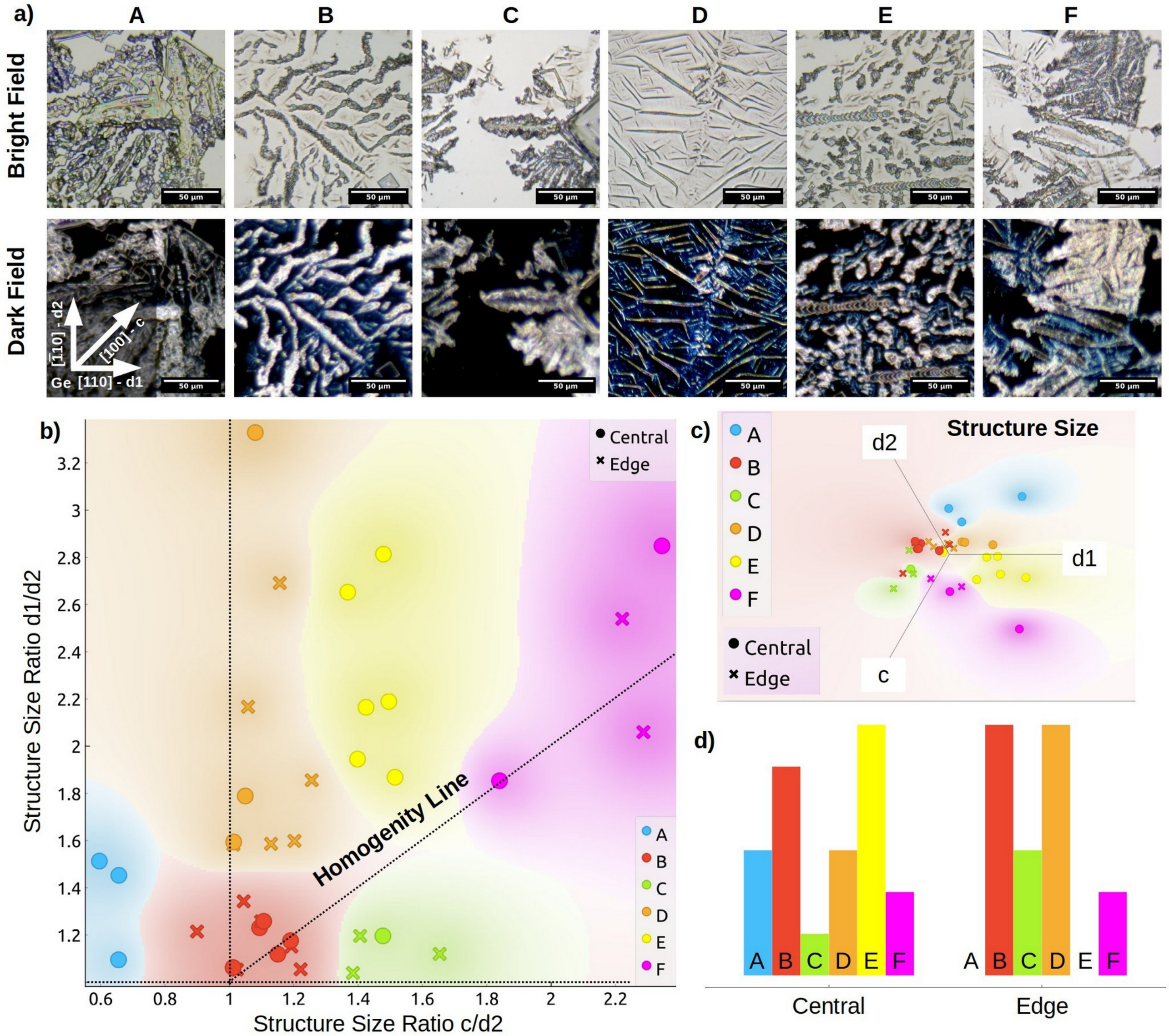

*Figure 2: Droplet deposition of NaCl on a Germanium (001) surface. a) Reflected light bright field and corresponding dark field microscopy images detailing the types of structures (A–F) formed after deposition. b) Structure size ratios (derived from autocorrelation analysis) along the directions d1/d2 versus c/d2 (where c is the average of c1 and c2 size) together with corresponding parallel coordinates plot (c-d2-d1) in c). Different regions on the plots correspond to the various structure types (A–F, indicated by different colors), which is a consequence of varying material diffusion strengths along different directions. We observe that structures formed along the indicated "homogeneity line" are homogeneously distributed throughout the droplet area. This line shows structures that possess similar diffusion strengths along the d1 and c directions. d) Histograms showing the occupancy of different structure types in the central part of the droplets versus the droplet edge. Structures A and F are observed exclusively in the central region.*

We also observe that structures formed along the indicated "homogeneity line", which represents regions where diffusion strengths are similar along the d1 and c directions, are uniformly distributed throughout the entire droplet area, structures B and F. This uniformity implies that under conditions of balanced diffusion along these two directions, the growth process proceeds in a consistent and predictable manner, leading to structurally coherent patterns without significant spatial inhomogeneity. The even distribution along this line further supports the idea that diffusion strength governs the final structure size and shape. Given that the growth originates from droplet deposition, the process is governed by a competition between two distinct physical mechanisms: the Marangoni effect and capillary flow. The Marangoni effect, driven by surface tension gradients, induces directed fluid motion away from regions of higher surface tension. This mechanism promotes lateral transport and can enhance diffusion along preferential directions, acting most strongly in the central region of the droplet. In contrast, capillary flow, governed by pressure gradients arising from the curvature of the liquid meniscus, drives redistribution of material towards the edges. As a result, it influences both the overall growth rate and the spatial uniformity of the structures, with its strongest contribution near the droplet edges.

The interplay between these two mechanisms modulates the effective diffusion field experienced by the growing structures and thereby contributes to the observed anisotropy in their size and morphology. Notably, the persistence of a well defined and uniformly populated “homogeneity line” suggests that, under particular conditions, the transport dynamics reach a quasi equilibrium in which Marangoni driven and capillary driven flows balance each other sufficiently to suppress spatial variations in growth dynamics. This dynamic balance stabilizes the effective diffusion field across the droplet, allowing the different structure types to be robustly correlated with diffusion strength.

We further observe that structure types A and E are exclusively present in the central region of the droplet after deposition, whereas structure types B and D are predominantly found near the droplet edges, see Fig. 2d). This spatial segregation is consistent with the position dependent dominance of Marangoni driven transport in the droplet interior and capillary driven redistribution near the contact line.

To obtain information about the formed thin film thickness, we performed a Quantitative Thin Film Interference analysis, as shown in Figure 3a). On the collected BF images, visible thin film interference fringes are observed. We compared the BF image with a model of NaCl thin film interference on germanium, as calculated using the approach by D. Lee and S. Lee[30]. The comparison was carried out pixel by pixel using least-squares minimization of the distance between the experimental image and the model in the CIE Lab color space, considering only the a and b values. Prior to the comparison, we corrected for illumination inhomogeneity in the BF images by subtracting the rolling ball background in ImageJ/Fiji[31]. After background subtraction, the germanium color was shifted to the theoretical germanium value. This resulted in the NaCl thin film thickness map. The quality of the fit is monitored pixel by pixel and is visually represented in the interpretable dE_CIELAB(a,b) error map and its corresponding histogram. Typically achieved dE_CIELAB values after fitting lie in the range of 1-2, which is colorimetrically considered a good result, further details are provided in Figure S4 of the supporting information.

Figure 3b) shows the thickness distribution of the exemplary NaCl film. The layer thickness is not continuous but exhibits five discrete peak values (M1 to M5), a behavior observed across all samples. The average position and width of the observed discrete film thickness peaks (M1 to M5) are presented in Table 1, with the calculated uncertainties reflecting the variability in peak position across all measured samples. It is evident that both the position and width of the peaks remain constant across all samples. For comparison, the average peak thickness positions for M3, M4, and M5 are presented as reported by C.W. Bunn and H. Emmett[32]. The observed peak positions in this work are consistent with previous measurements. We also see that the thickness peak positions are consistent with quantization model L*n*(n+1), where L=12.93±0.14 nm is effective relaxation length, see Supporting Information Figure S12.

| | M1 | M2 | M3 | M4 | M5 |
|---|---|---|---|---|---|
| **This Work**<br>Average Position of peak Thickness [nm] | 33.45±0.95 | 82.5±1.3 | 162.8±3.6 | 266.4±3.6 | 371.5±3.9 |
| **This Work**<br>Width (Std.Dev.) of peak Thickness [nm] | 5.52±0.49 | 5.7±1.7 | 13.1±1.5 | 9.9±3.6 | 16.2±5.2 |
| C.W. Bunn and H. Emmett[32]<br>Average Position of peak Thickness [nm] | - | - | 170 | 280 | 350/360 |
| **This Work**<br>Quantization Model Thickness [nm]<br>L*n*(n+1), L=12.93±0.14 nm | 25.86±0.28 | 77.58±0.84 | 155.2±1.7 | 258.6±2.8 | 387.9±4.2 |

*Table 1: Droplet deposition of NaCl thin film on a Germanium (001) surface. The average position and width of the observed discrete film thickness peaks (M1 to M5) are shown, with the calculated uncertainties reflecting the variability in peak position across all measured samples. It is evident that both the position and width of the peaks remain constant across all samples. For comparison, the average peak thickness positions for M3, M4, and M5 are presented as reported by C.W. Bunn and H. Emmett[32]. The observed peak positions in this work are consistent with previous measurements. The proposed thickness quantization model L*n*(n+1), where L=12.93±0.14 nm is effective relaxation length, is also presented.*

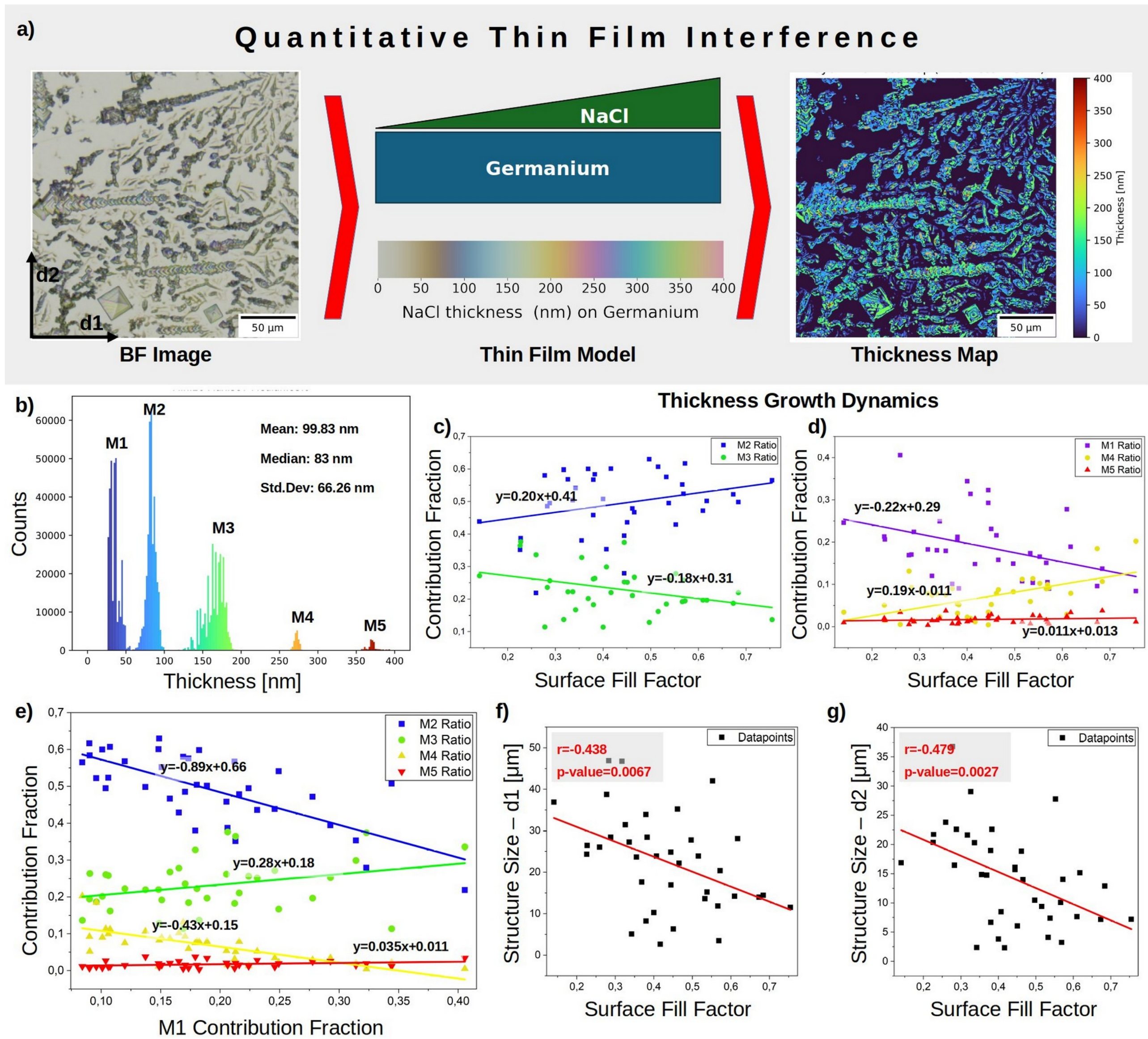

*Figure 3: Results of Thin Film Thickness Analysis from Bright Field Microscopy of NaCl Droplet Deposition on a Germanium (001) Surface. a) Schematic illustrating the Quantitative Thin Film Interference analysis: the experimental bright field image is fitted to a thin film interference model of NaCl on Germanium to generate a thickness map of the NaCl film. b) Thickness distribution of the exemplary NaCl film. The layer thickness is not continuous but exhibits five discrete values (M1 to M5), a behavior observed across all samples. c) and d) Contribution fractions of M1, M3, and M1, M4, M5 maxima, respectively, as a function of surface filling factor, along with linear function fits. The contribution fraction is correlated with the surface filling factor for all maxima, and the linear rate 'a' is similar across all, indicating coupled growth dynamics for different thicknesses. e) Thickness contribution fractions of M2, M3, M4, and M5 as a function of the M1 contribution. These fractions show a linear correlation between each other, further supporting coupled growth. f) and g) Structure size (from autocorrelation analysis) as a function of surface filling factor for the d1 and d1 directions. Structure size is inversely correlated with the surface filling factor. These results are statistically significant ($p<0.05$), further supporting surface-limited growth dynamics.*

Figure 3c)–d) shows the contribution fractions of each M1, M3 and M2, M4, M5 thickness peaks to the total film thickness, normalized to sum to 1, respectively, as a function of surface filling factor,

along with linear function fits (y = ax + b). It is seen that the contribution fraction is correlated with the surface filling factor for all thickness peaks, and the linear rate 'a' is similar across all, indicating coupled growth dynamics for different thicknesses. Figure 3e) shows the thickness contribution fractions of M2, M3, M4, and M5 as a function of the M1 contribution. These fractions exhibit a linear correlation between each other, further supporting coupled growth dynamics. Figure 3g) shows structure size as a function of surface filling factor for the d1 and d1 directions. It is seen that structure size is inversely correlated with the surface filling factor. These results are statistically significant ($p < 0.05$), further supporting surface-limited growth dynamics. These observations reveal a linked growth mechanism in which the formation of discrete thickness peaks (M1 to M5) is not governed by independent or stochastic processes, but rather by a common, surface-mediated dynamics. The parallel evolution of peak contributions with surface filling factor, the linear dependence among the M2-M5 fractions relative to M1, and the inverse scaling of structure size with coverage all point to a common physical origin of surface-limited growth under constrained interfacial conditions. The consistent slopes of the linear fits across different peaks suggest that the system responds uniformly to changes in surface occupancy, implying a shared control parameter, such as surface energy, local diffusion rates, or wetting-induced capillary forces, that governs the nucleation and stability of each thickness peak. This coupled behavior indicates that the discrete peaks are not merely statistical artifacts, but represent distinct, dynamically stabilized morphological states that emerge from a collective balance between deposition, surface interactions, and spatial constraints. Furthermore, the inverse relationship between structure size and filling factor supports the idea that growth is fundamentally limited by the availability of active surface sites, with larger structures forming only when sufficient space and low local density are available. These findings support a model in which thin film morphology evolves not through random or independent layering events, but through a self-organizing, surface-constrained process where thickness distribution, peak contributions, and feature size are linked. Notably, we do not see any correlations between layer thickness and structure size, suggesting that the growth processes in the plane and out of plane are decoupled.

As we showed, the proposed Quantitative Thin Film Interference analysis successfully enables the description and characterization of thin film thickness growth. Additionally, we performed evaluation of the method to account for the effects of defocus and image noise, as in real experiments. These factors were included using simulated bright-field (BF) images of NaCl thin film on germanium, as shown in Figure S5 and Table S1 of the Supporting Information. From this study, the corresponding relative maximal error related to defocus and noise in the thickness determination via the proposed method was estimated to be 3.5%. It is also worth noting that the film thickness determined by the method is consistent with SEM cross-sectional tilt view measurements, as shown in Figure S6 of the Supporting Information.

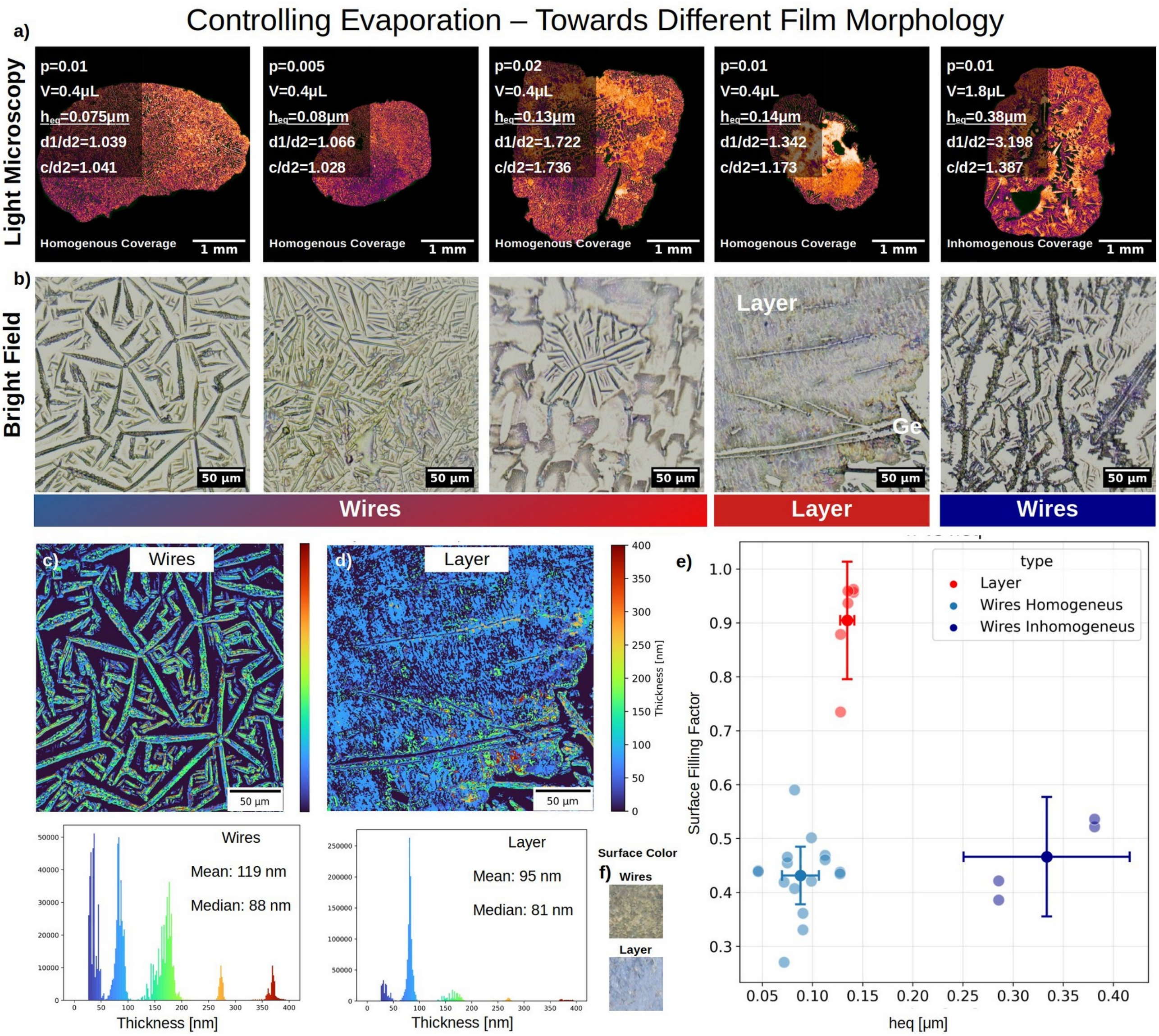

*Figure 4: Volume and concentration dependence of NaCl droplet deposition on a Ge(001) surface. a) and b) Exemplary light microscopy and bright-field microscopy images for different droplet volumes and concentrations, illustrating varying evaporation dynamics and resulting equivalent thicknesses ($h_{eq}$). Deposition parameter indicated. The morphology evolves systematically from wire structures to compact layers as $h_{eq}$ increases. c) and d) Thickness maps and corresponding histograms for homogeneous wires and compact layer, showing a well-defined, uniform thickness for the layer. e) Surface filling factor as a function of $h_{eq}$ (NaCl equivalent thickness). As $h_{eq}$ increases from 0.075 µm to 0.14 µm, the morphology transitions from homogeneous wires (filling factor ~0.4) to a compact layer (filling factor ~0.9). For $h_{eq}$ values above 0.14 µm, inhomogeneous wire structures appear. f) surface color of the homogeneous wires (yellow tint) sample and layer sample (blue tint) as seen by Light Microscopy. It is worth to notice that all other samples type appear white.*

This together confirms the reliability and practical applicability of the proposed Quantitative Thin Film Interference analysis as a valuable, accessible tool for thin film thickness characterization in real experimental conditions. The Python code implementing the Quantitative Thin Film Interference analysis, together with the NaCl on germanium thickness model, is freely available in the Zenodo repository[33]. In order to access different $h_{eq}$ values and distinct evaporation dynamics, we performed additional experiments for various droplet volumes and NaCl concentrations, as shown in Figure 4. Fig. 4a)–b) shows LM images together with detailed views as obtained in bright field microscopy. The morphology evolves systematically from homogeneous wire structures in the whole droplet area to compact homogeneous layer as $h_{eq}$ increases, before reverting back to inhomogeneous wire-like structures. The thickness analysis shows that the homogeneous wires exhibit a range of different thicknesses, while the compact layer exhibits almost one well defined thickness, see Fig. 4c)–d). The detailed study of the surface filling factor (from BF images) as a function of $h_{eq}$, see Fig. 4e), shows that as $h_{eq}$ increases from 0.075 μm to 0.14 μm, the morphology transitions from homogeneous wires (filling factor ~0.4) to a compact layer (filling factor ~0.9). For $h_{eq}$ values above 0.14 μm, inhomogeneous wire structures appear. It is seen that the homogeneous layer is formed only within a very specific range of $h_{eq}$ values. This observation highlights the delicate balance between evaporation dynamics and solute deposition kinetics in NaCl droplet systems. Since $h_{eq}$ represents the equivalent height of pure NaCl, it serves like a proxy for the effective concentration and evaporation rate of the salt solution during droplet drying. The evaporation time is proportional to the droplet radius[34]. Within the narrow $h_{eq}$ range where homogeneous wire structures are observed, the evaporation time is sufficiently large to allow for uniform ion distribution and consistent nucleation, enabling the formation of uniform, wire morphologies. As $h_{eq}$ increases beyond this window, the evaporation time accelerates, leading to faster depletion of the solution and a shift towards localized supersaturation. This promotes rapid and non-uniform nucleation, resulting in the formation of compact layers due to enhanced surface coverage and aggregation of NaCl crystals. Conversely, when $h_{eq}$ exceeds 0.14 μm, the evaporation becomes too rapid, causing the solution to dry before complete redistribution of ions, which in turn leads to inhomogeneous nucleation and the re-emergence of wire-like structures but in this case inhomogeneous. This emphasizes the critical role of evaporation kinetics in determining the final morphology during NaCl deposition from droplets. Homogeneous wires arise transiently due to balanced evaporation, diffusion, and supersaturation, precise control of $h_{eq}$ is key for reproducible morphologies. One can also see that the homogeneous wires as well as the layer structure lie on the "homogeneity line" (with similar size ratios d1/d2 and c/d2), where the transport dynamics reach a quasi-equilibrium. It is also worth noting that homogeneous wires exhibit a distinct yellow color tint, while the layer shows a blue color tint under Light Microscopy (diffuse light), all other samples appear white in color, see Fig. 4f). Suggesting appearance of some optical effects.

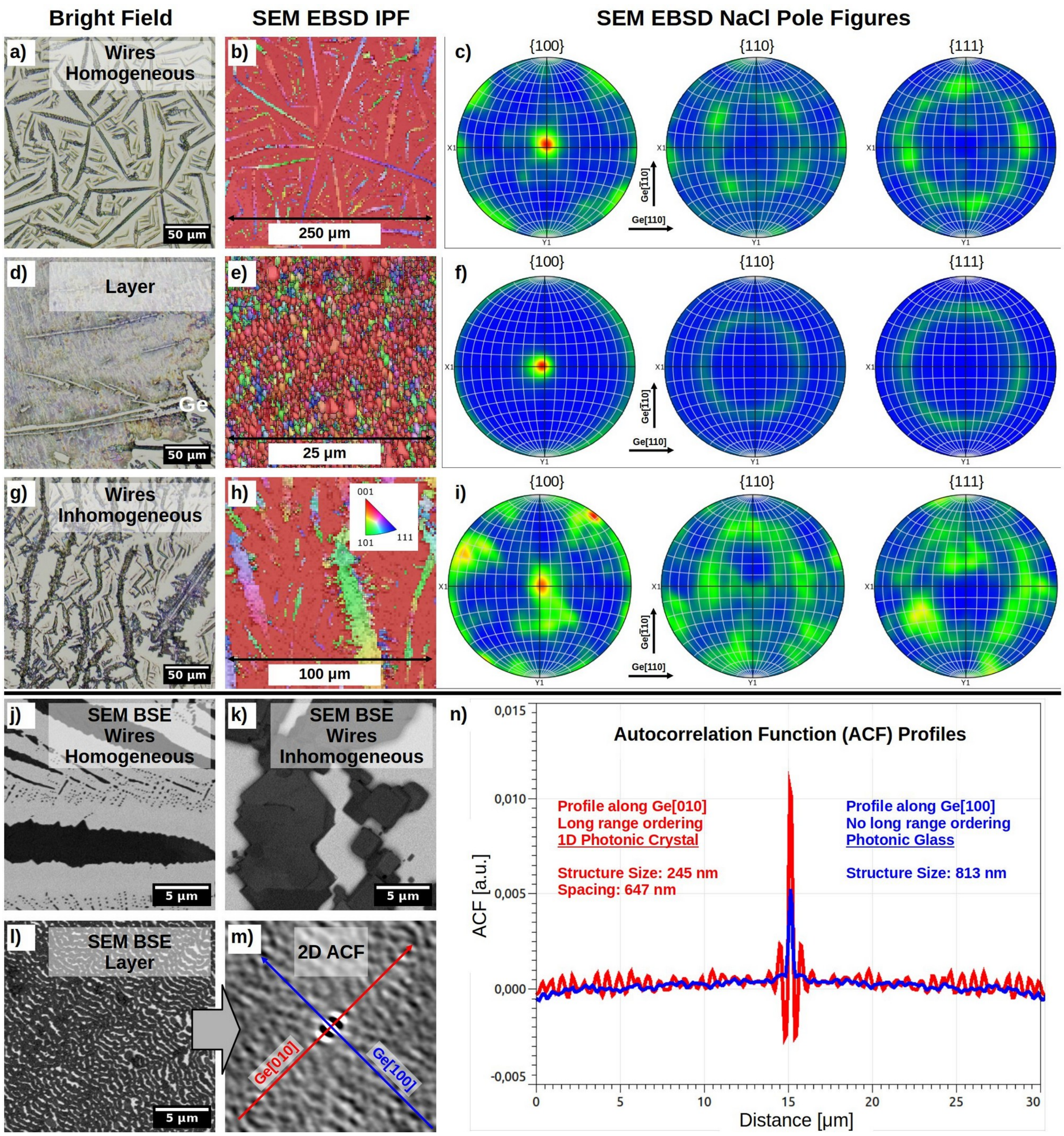

*Figure 5: SEM EBSD analysis of NaCl droplet deposition on a Ge(001) surface. a), d), g) Bright-field microscopy images together with corresponding EBSD IPF maps b), e), h) and EBSD pole figures c), f), i) for homogeneous wires, compact layer, and inhomogeneous wires, respectively. In all cases NaCl (001) layers grow epitaxially on the Ge(001) surface, exhibiting different in-plane rotational symmetries. j), k), l) SEM BSE images reveal detailed morphology: homogeneous wires, inhomogeneous wires, and the compact layer composed of small, densely packed nanocrystallites. m) 2D autocorrelation function (ACF) of the compact layer, showing long-range order along the Ge[010] direction. n) ACF profiles along Ge[010] and Ge[100] directions: long-range nanocrystallite ordering is observed along Ge[010], indicating the formation of a 1D photonic crystal. No such ordering is found along Ge[100], suggesting a photonic glass phase. The resulting compact NaCl layer thus exhibits hybrid (crystal–glass) photonic behavior.*

To obtain detailed information about the crystallography and morphology of the formed homogeneous structures within the droplet area, we performed SEM (Scanning Electron Microscopy) and EBSD (Electron Backscatter Diffraction) measurements, see Figure 5. The EBSD IPF maps and corresponding EBSD Pole Figures (PF), see Fig. 5a)-i), show that for all cases NaCl (001) layers grow epitaxially on the Ge(001) surface, with {100} PFs clearly showing the pole. It is also evident, mainly from the {110} PFs, that the structures exhibit different in-plane rotational symmetries. For the homogeneous wires, particularly enhanced regions of specific angles are observed, which evolve into perfect homogeneous rotations for the layers (manifesting as a perfect circle), and finally transition to random rotations for the inhomogeneous wires. The EBSD results show that the degree of structural homogeneity is directly linked to the consistency of growth dynamics and interfacial stability. Homogeneous structures achieve a high degree of rotational order due to a balanced and stable growth environment, while inhomogeneous structures suffer from disordering effects that degrade crystallographic uniformity. This highlights the critical role of growth conditions in determining the final crystal morphology and orientation, with the "homogeneity line" serving as a key regime where such ordered growth is effectively maintained. This is also reflected in the appearance and evolution of different in-plane orientations, as shown in the Inverse Pole Figures (IPF) Fig. S7–S9 in the Supporting Information. We also observe that the larger NaCl crystallites (~20 microns) exhibit a particular and consistent in-plane rotation across the samples, with respect to the germanium crystallographic directions. The NaCl crystallites are consistently rotated by 10.11 degrees, corresponding to their orientation <9013> relative to Ge <101>, as shown in Fig. S10 of the Supporting Information. We believe this effect is purely related to energy minimization of the system during growth, particularly due to Coulomb interactions. On the reconstructed germanium (001) surface, atomic wires along the <110> directions exist, so if the NaCl alignment was perfect, Na or Cl atoms would align with these wires, bringing charged ions (both negative or both positive) into close proximity within the wire-like structure. This rotated alignment minimizes the overall system energy by maximizing charge screening and promoting favorable ion-surface interactions along the wire channels. For the smaller crystallites, however, this effect is partially offset by local configurational adjustments, small-scale distortions or reconstructions that allow for a more flexible accommodation of the ionic lattice relative to the underlying substrate. These local modifications are sufficient to mitigate the strong energetic penalty associated with perfect alignment, as demonstrated by DFT calculations[35], which reveal a reduction in interfacial strain and a stabilization of the ionic arrangement through subtle atomic shifts. While the large crystallites maintain a robust, uniform orientation driven by long-range energetic optimization, the smaller ones exhibit greater structural variability, reflecting the influence of size-dependent surface effects and limited spatial extent for charge screening. This contrast underscores the importance of both scale and interfacial energetics in governing the orientation and stability of NaCl grown on Ge surfaces.

SEM BSE imaging, where the contrast is proportional to atomic number and thickness, reveals detailed features of the sample morphology, as shown in Fig. 5j)–l). Particularly, it is revealed that the compact layer is actually composed of small, densely packed NaCl nanocrystallites, as evident in Fig. 5l) and Fig. S6 in the Supporting Information. The performed 2D ACF (autocorrelation function) analysis of the layer image uncovers the nanocrystallites ordering, as clearly demonstrated in Fig. 5m). The ACF profiles along the Ge[010] and Ge[100] directions reveal the spatial correlation characteristics of the crystallite arrangement, providing detailed insight into the degree and directionality of structural ordering, as shown in Fig. 5n). Long-range ordering is observed exclusively along the Ge[010] direction, with a well-defined structure size of 245nm and a characteristic spacing of 647nm, these parameters are consistent with periodic, repeatable packing of nanocrystallites, indicating the formation of a one-dimensional photonic crystal[36]. In contrast, no such long-range order is detected along the perpendicular Ge[100] direction and other ones, where the structure size is significantly larger at 813nm and exhibits broad, decaying correlation peak, this behavior is characteristic of a disordered, amorphous-like arrangement, consistent with the formation of a photonic glass phase. It is worth to notice that we see similar 2D ACF behavior for high magnification BF and BF images, see Fig. S11 in Supporting Information. As a result, the compact NaCl layer exhibits a hybrid photonic behavior, combining the well-defined, periodic light-matter interactions of a 1D photonic crystal along Ge[010] with the diffuse, broadband scattering properties typical of a photonic glass along other directions. This dual nature implies that the layer supports both guided wave propagation along the ordered Ge[010] direction and broad spectral interference effects arising from the disordered phase, which could be exploited in applications requiring tunable optical responses or multi-directional light manipulation, particularly in the infrared (IR) range. Since both germanium and sodium chloride exhibit strong transparency and favorable optical properties in the mid to far infrared, their compatibility in this spectral region might enable efficient light confinement and interaction with the structured nanocrystalline interface.

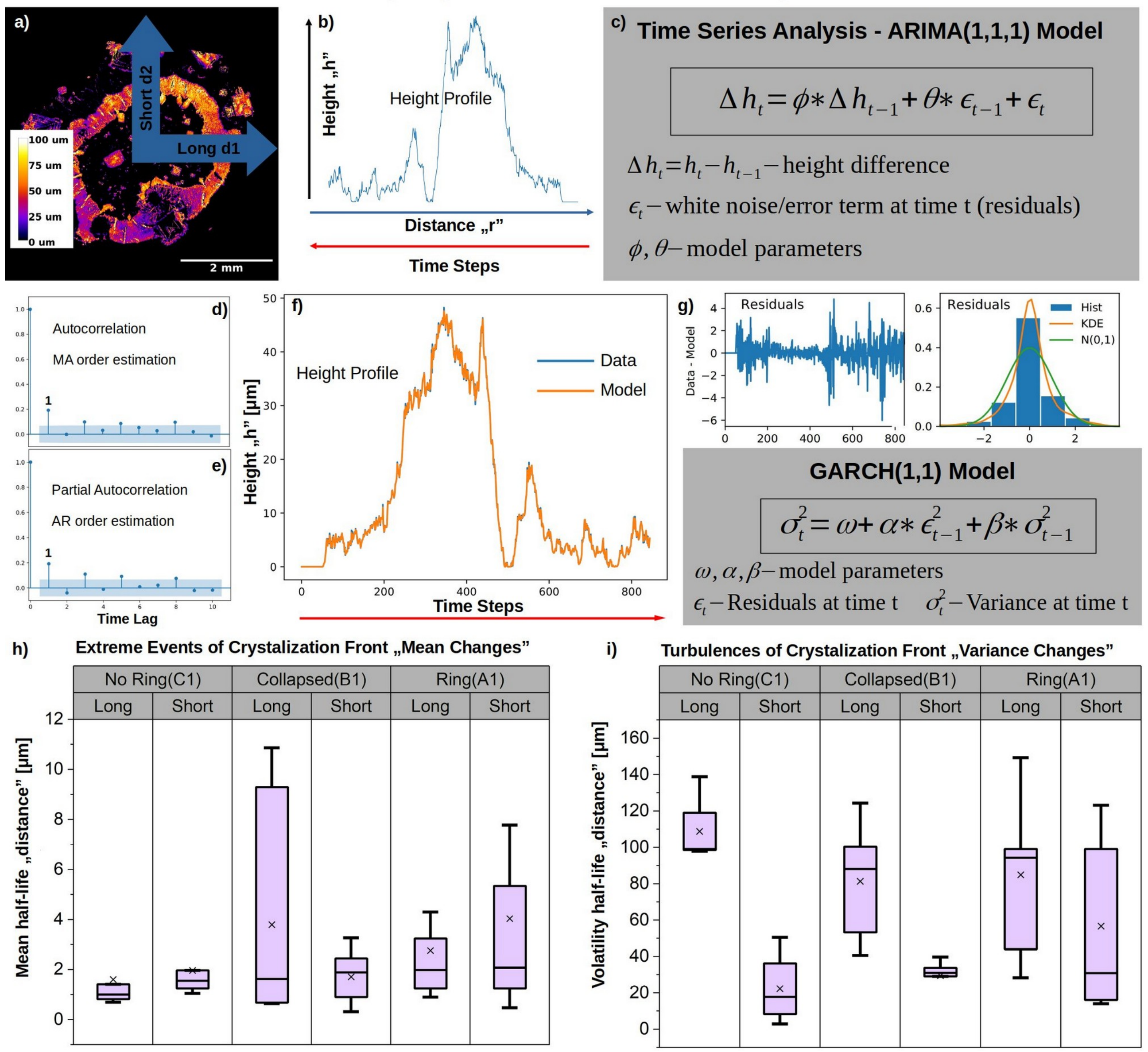

*Figure 6: Time series analysis of crystallization front dynamics in NaCl droplet deposition on a Ge(001) surface. a) Height map from a light microscopy image with profiles along directions d1 and d2. b) Exemplary height profile, where increasing distance corresponds to decreasing crystallization time. c) Fitted ARIMA(1,1,1) model in the time domain (reverse distance). d) Autocorrelation function and e) partial autocorrelation function, both showing a peak at lag 1, indicating non-random, structured dynamics. f) Exemplary height profile as a function of time steps (reverse distance in pixels), along with the fitted ARIMA model. g) Residuals (data - model) versus time steps, showing non-Gaussian distribution, suggesting heterogeneity in the process. A GARCH(1,1) model is applied to describe the residual volatility dynamics. Final results: h) Box plots of mean half-life for extreme events, illustrating the dynamics of mean height changes, i) Box plots of volatility half-life (i.e., time scale for variance/roughness fluctuations), showing that mean half-lifes are on the order of a few microns, while volatility half-lifes are several hundred microns. Systematic directional dependencies are observed along long (d1) and short (d2) directions, as well as in cases where coffee ring effects appear.*

The formed structures are a consequence of droplet deposition, where during evaporation the crystallization front moves from the edge of the droplet towards the center. The resulting crystalline structures and their spatial distribution (i.e., height distribution) along the droplet area are a direct result of the dynamics of the moving crystallization front, such that the edge corresponds to the initial time and the center to the final time of crystallization events. Thus, the height profile from center to edge can be treated as a reverse time series of the crystallization events, see Figure 6a)–b). To describe this time series, we applied the full machinery of well-known time series analysis, as commonly used in financial modeling[37]. We performed an ARIMA[38] model fit, AutoRegressive Integrated Moving Average, a statistical modeling technique used to understand and forecast time series data based on its own past values, differences, and error terms. The ARIMA(1,1,1) model was fitted to the final droplet height profiles after deposition along the d1 and d2 directions for each sample. The model equation is given by: $\Delta h_t = \phi * \Delta h_{t-1} + \theta * \epsilon_{t-1} + \epsilon_t$ , where $\Delta h_t$ is the height difference, $\epsilon_t$ is the white noise/error term (residuals), $\phi$ and $\theta$ are model parameters, and t denotes the time step in our case in pixels, which can later be translated into physical distance. Before fitting the time series model, a Box-Cox transformation[39] (a standard procedure) was applied to ensure stationarity of the height profile time series. Figure 6d) shows the autocorrelation function and Figure 6e) the partial autocorrelation function, both of which exhibit clear peaks at lag 1. A value of zero at lag 1 would indicate a random walk. The observed peaks confirm that the dynamics are non-random and structured. This also clearly validates the presence of both AR (autoregressive) and MA (moving average) components in the ARIMA model. Figure 6f) shows an exemplary height profile as a function of time steps (reverse distance in pixels), along with the fitted ARIMA model. It is evident that the model captures the data very well. A detailed analysis of the residuals (data-model) versus time steps, shown in Figure 6g), reveals a non-Gaussian distribution, suggesting heterogeneity in the underlying process. To further describe this heterogeneity, we additionally fitted a GARCH[40] model (Generalized Autoregressive Conditional Heteroskedasticity) to the residuals, a statistical framework widely used in econometrics and financial analysis to model and forecast the volatility of asset returns and to capture the dynamics of residual volatility. We successfully fitted the following GARCH(1,1) model equation: $\sigma_t^2 = \omega + \alpha * \epsilon_{t-1}^2 + \beta * \sigma_{t-1}^2$ , where $\omega, \alpha, \beta$ are model parameters, $\epsilon_t$ are the residuals, $\sigma^2_t$ is the conditional variance, and t denotes the time step. From the ARIMA and GARCH model fits, physically meaningful quantities can be derived, specifically, half-lifes to quantitatively describe the dynamics of the crystallization front. The mean half-life for extreme height events was calculated from the ARIMA model using the equation $\ln(0.5)/\ln(\phi)$, indicating the time after which extreme height spikes relax by 50% (i.e., a reduction in height by half), thereby illustrating the dynamics of mean height changes. Similarly, from the GARCH model, the volatility half-life is determined via ln(0.5)/(α+β), representing the time after which extreme fluctuations in height variance (i.e., RMS roughness) reduce by 50%, thus revealing the time scale of roughness fluctuations and turbulence. The results are presented in Figure 6h) and Figure 6i) as box plots for different directions (d1 - Long, d2 - Short) and for different groups according to coffee ring appearance (different $h_{eq}$ values). Immediately, it is seen that the mean half-life for height relaxation is significantly lower, in the order of a few microns, compared to the volatility half-life associated with roughness (~hundreds of microns). This demonstrates that height spikes during crystallization relax much faster than roughness spikes (turbulences). This observation has important physical implications: it suggests that the dynamics governing the local growth of crystal nuclei are rapid and highly localized, driven by short-range diffusion and surface energy minimization. In contrast, the evolution of surface roughness is governed by longer-range interactions, such as lateral flow instabilities, interfacial tension gradients, and stochastic fluctuations in evaporation rates. These processes accumulate over larger spatial scales and exhibit slower temporal relaxation, leading to the persistence of roughness features even after the peak height events have subsided. Thus, the faster decay of height spikes compared to roughness fluctuations indicates a decoupling between the kinetics of crystal nucleation and the macroscopic surface morphology evolution. Therefore, the separation of time scales, with height relaxation occurring on micrometer scales and roughness dynamics persisting over hundreds of microns reveals a hierarchical structure in the system i.e. fast, localized events drive immediate changes in crystal height, while slower, large-scale instabilities govern the long-term surface texture. Additionally, the appearance of a coffee ring significantly increases the mean half-life. This is a direct consequence of the enhanced transport of solutes toward the droplet periphery due to the evaporation-driven capillary flow. As the coffee ring effect develops, solute concentration builds up at the edges, promoting more crystallization along the rim. This results in a broader, more stable front that propagates inward more slowly, thereby extending the time required for height relaxation. The increased half-life reflects a shift from a transient, edge-dominated crystallization process to a more persistent, rim-anchored growth mechanism. Regarding volatility half-life, we observe that as the system transitions from no coffee ring to coffee ring formation, the volatility half-life decreases in the d1 (long) direction while it increases in the d2 (short) direction. This directional asymmetry provides critical insight into the anisotropy of evaporation and solute transport. The reduction in volatility half-life along d1 (long) suggests that the flow in this direction is more efficiently coupled to the surface dynamics, likely due to a stronger material diffusion along that axis. As a result, fluctuations in height variance are damped more rapidly, indicating a more stable and predictable surface evolution in this direction. Conversely, the increase in volatility half-life along d2 (short) implies that the transport in this direction is less efficient, leading to prolonged and irregular roughness fluctuations. In all cases, the volatility half-life is always higher in the long direction (d1) than in the short direction (d2). This consistent asymmetry points to an underlying anisotropic transport field, where the evaporation rate and solute diffusion are not uniform across the droplet. Crucially, d1 and d2 correspond to crystallographic <110> directions on germanium with atomic chains wires which act as low-resistance pathways for fast and efficient solute diffusion. We believe that the d1 direction aligns with the long terrace edge, where these atomic wires extend continuously and provide a preferential channel for ion transport. As a result, solutes diffuse more rapidly along d1, leading to faster surface equilibration and more homogeneous nucleation. This explains the observed faster damping of height fluctuations, the system reaches a dynamic steady state more quickly due to enhanced transport. In contrast, the d2 direction, which lies perpendicular to the terrace edge i.e. shorter atomic wire path, exhibits more scattering and resistance to diffusion, this results in slower transport.

Thus, the performed ARIMA and GARCH model fits to the height profiles do not merely describe data, they reveal the intrinsic time-dependent physics of the droplet crystallization process under evaporation. By extracting half-lifes from these models, we move beyond descriptive statistics to quantify the dynamics of surface evolution in a way that is directly linked to physical mechanisms. These metrics provide a robust, quantitative language to compare different deposition conditions, materials, or environmental settings, enabling a deeper understanding of how external factors (such as evaporation rate, droplet volume, solute concentration) influence the stability and structure of the final crystalline pattern. In this sense, the time series analysis becomes not just a mathematical tool, but a window into the underlying physical processes that shape the final morphology of the formed crystalline structures.

## Summary and Conclusions

We performed a comprehensive, multi-scale investigation into the crystallization dynamics of sodium chloride (NaCl) during droplet evaporation on a germanium (001) substrate surface. By integrating microscopy characterization with quantitative thin-film analysis and time-series modeling, we uncover the mechanisms governing the emergence of complex crystalline morphologies, ranging from coffee-ring patterns to hybrid photonic phases. A central element of our approach is the introduction of the NaCl equivalent height ($h_{eq}$), which serves as a unified parameter capturing deposition coverage and enabling systematic comparison of different experimental conditions. We observe that decreasing $h_{eq}$ leads to a reduction in both film height and roughness, while structural features exhibit a characteristic growth and saturation behavior. Autocorrelation analysis reveals a pronounced anisotropy in structure size along crystallographic directions, indicating that surface diffusion is strongly directional. The comparable growth length scales across directions suggest that diffusion anisotropy primarily determines the maximal attainable structure size, while the underlying growth kinetics remain similar. This diffusion-driven behavior gives rise to distinct morphological regimes that are governed by the balance of transport processes during evaporation. In this context, we identify a "homogeneity line," corresponding to conditions where diffusion is effectively balanced across crystallographic directions. Along this line, spatially uniform and reproducible structures emerge, reflecting a dynamic equilibrium between Marangoni-driven flows and capillary transport. This balance defines a key organizing principle for morphology selection during droplet evaporation. The thin-film thickness, as studied by our Quantitative thin film interference analysis approach, further reveals discrete, well-defined thickness states, which remain consistent across all samples. The absence of correlation between film thickness and lateral structure size demonstrates a clear decoupling between in-plane and out-of-plane growth dynamics. Instead of independent nucleation events, the system evolves through a coupled, surface-limited process governed by shared interfacial parameters, indicating a self-organizing mechanism that controls thickness selection. By varying droplet volume and concentration, we observe systematic transitions between homogeneous wire-like structures, compact layers, and inhomogeneous morphologies. Notably, all homogeneous states lie on the homogeneity line, reinforcing the central role of balanced transport dynamics in determining structural stability and reproducibility. Structural homogeneity is further reflected in well-defined crystallographic orientations, while deviations from this balance lead to disorder and variability. At the nanoscale, the compact layer consists of densely packed nanocrystallites that exhibit long-range order along one crystallographic direction, characteristic of a one-dimensional photonic crystal, while remaining disordered in perpendicular directions. This coexistence of ordered and disordered arrangements results in a hybrid photonic (crystal–glass) phase, offering potential for directional and broadband light manipulation, particularly in the infrared regime. Finally, the dynamics of the crystallization front can be described within a time-series framework, ARIMA and GARCH, revealing a clear separation of time scales, as seen with extracted half-lifes. Fast, localized relaxation of height fluctuations is governed by short-range diffusion and interfacial energetics, whereas slower evolution of surface roughness reflects large-scale transport processes and evaporation-induced instabilities. This hierarchy establishes a direct link between statistical dynamics and the underlying physical mechanisms that shape the final morphology.

Overall, our work establishes a unified framework for understanding evaporative crystallization across multiple scales. By linking morphology, transport dynamics, and interfacial physics through a single governing parameter ($h_{eq}$) together with the concept of the homogeneity line and time-series dynamics, we provide both a predictive description, physical understanding and a practical pathway for controlling structure formation. These findings open new opportunities for the controlled design of functional thin films and hybrid photonic systems through precise experiments of evaporation-driven processes.

**Acknowledgments**

This research was supported in part by the Excellence Initiative - Research University Program at the Jagiellonian University in Krakow. This work is the result of Grzegorz S. Żmija master thesis project.

## Author Contributions

G.S.Ż. planned and performed all deposition experiments together with microscopy measurements LM/BF/DF and data analysis. G.C. contributed to SEM EBSD measurements and data analysis. B.R.J. implemented Quantitative Thin Film Interference Analysis in python. B.R.J. conceived the idea, supervised and organized the project. B.R.J. prepared the manuscript in consultation with all authors. All authors contributed to the discussion and interpretation of the final results.

**Conflicts of Interest**

The authors declare no conflicts of interest.

**Data Availability**

All the experimental collected data together with python program code of Quantitative Thin Film Interference Analysis are freely available in Zenodo repository https://doi.org/10.5281/zenodo.20085570.

# Supporting Information

## The Emergence of Photonic Crystalline Order and Time-Series Dynamics in NaCl Droplet Deposition

Grzegorz S. Żmija[1], Grzegorz Cios[2], Benedykt R. Jany[1*]

[1]Marian Smoluchowski Institute of Physics, Faculty of Physics, Astronomy and Applied Computer Science, Jagiellonian University, ul. prof. Stanisława Lojasiewicza 11, 30-348 Krakow, Poland
[2]Academic Centre for Materials and Nanotechnology, AGH University of Krakow, al. A Mickiewicza 30, 30-059 Krakow, Poland

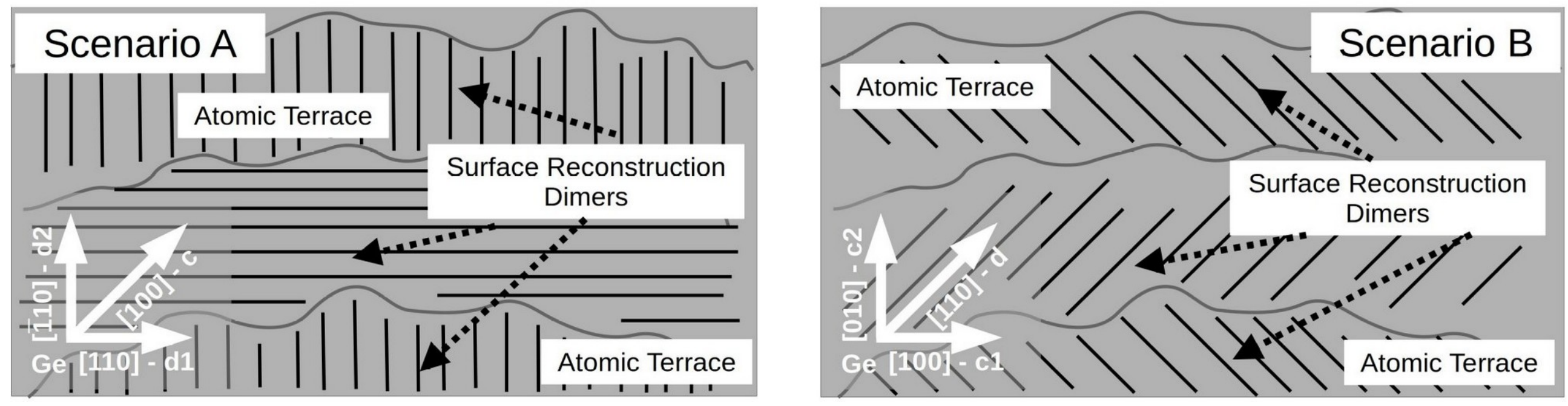

*Figure S1: Schematic illustrating two scenarios of reconstructed Ge(001) surfaces: Scenario A and Scenario B. Crystallographic directions d1 (Long), d2 (Short), c1, and c2 are indicated.*

* corresponding author e-mail: benedykt.jany@uj.edu.pl

**Height maps from Light Microscopy Images**

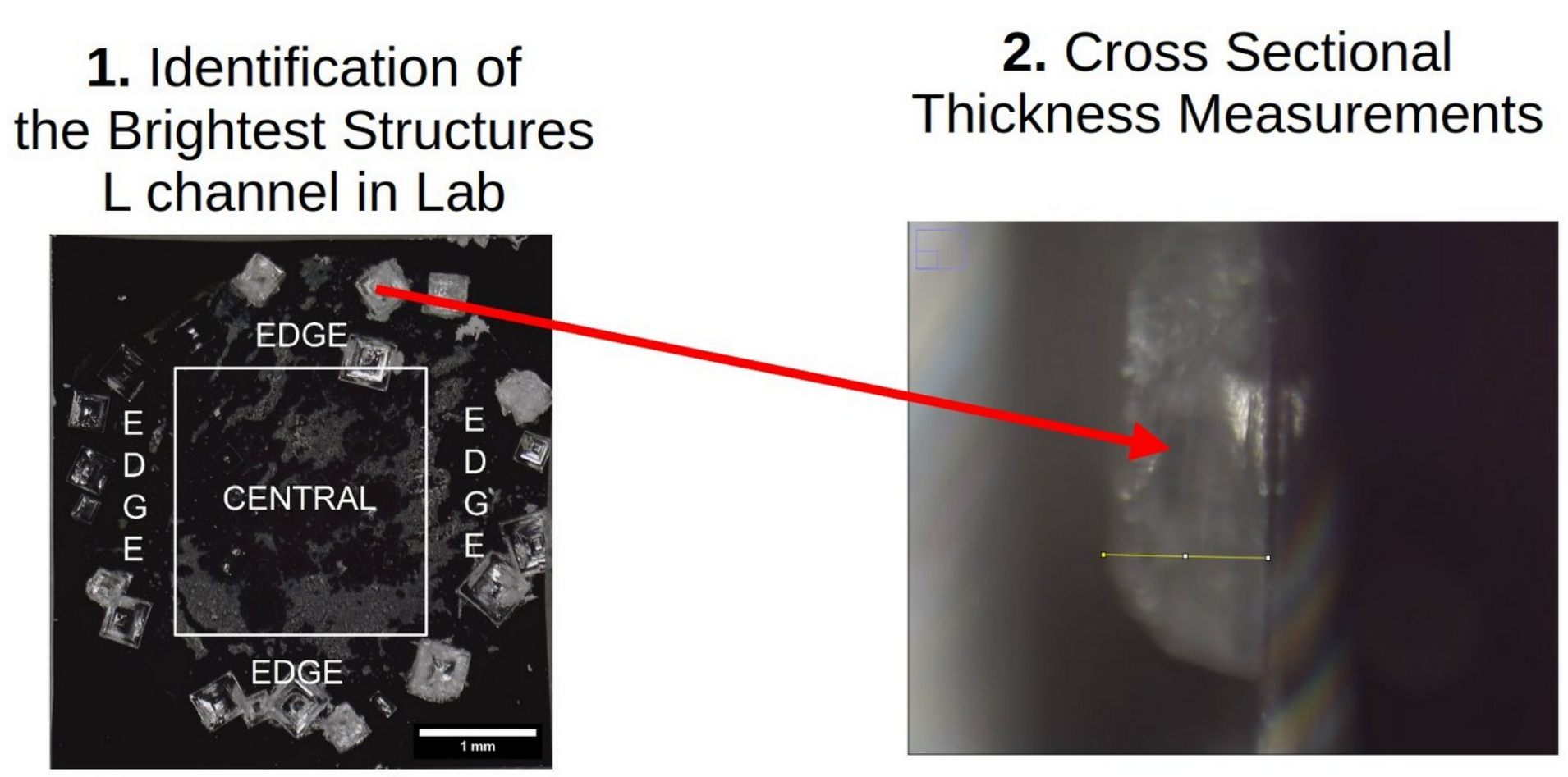

**3.** Inverted Lambert-Beer Model
Transparent salt masking dark germanium

$$h = \begin{cases} L \cdot 0,00689 - 0,0422; & L < 22,5 \\ -36,76654 * \ln(\frac{(66.4855-L)}{(66,4855-16,16983)}); & 22,5 < L < 66,45 \\ 0; & L \geq 66,45 \end{cases}$$

Datapoints
ThinFilm Regime
Thick Regime
height [µm]
Intensity

**4.** Recalculation of the L channel
into height Map

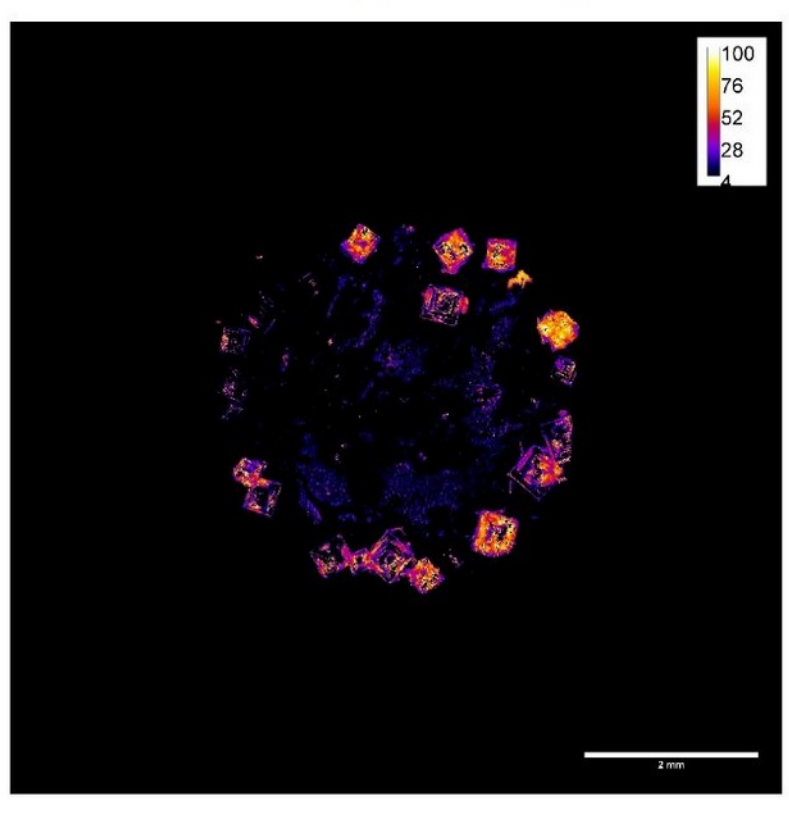

*Figure S2: Schematic showing idea of the height map generation from Light Microscopy images via cross sectional height measurements and effective inverted Lambert-Beer model parametrization. We used piecewise function to account also for the thin film region in additionally to the micro crystallites. Finally effective data based parametrization which translate lightness into height was derived.*

### XRD Measurements of NaCl on Ge(001) surface samples

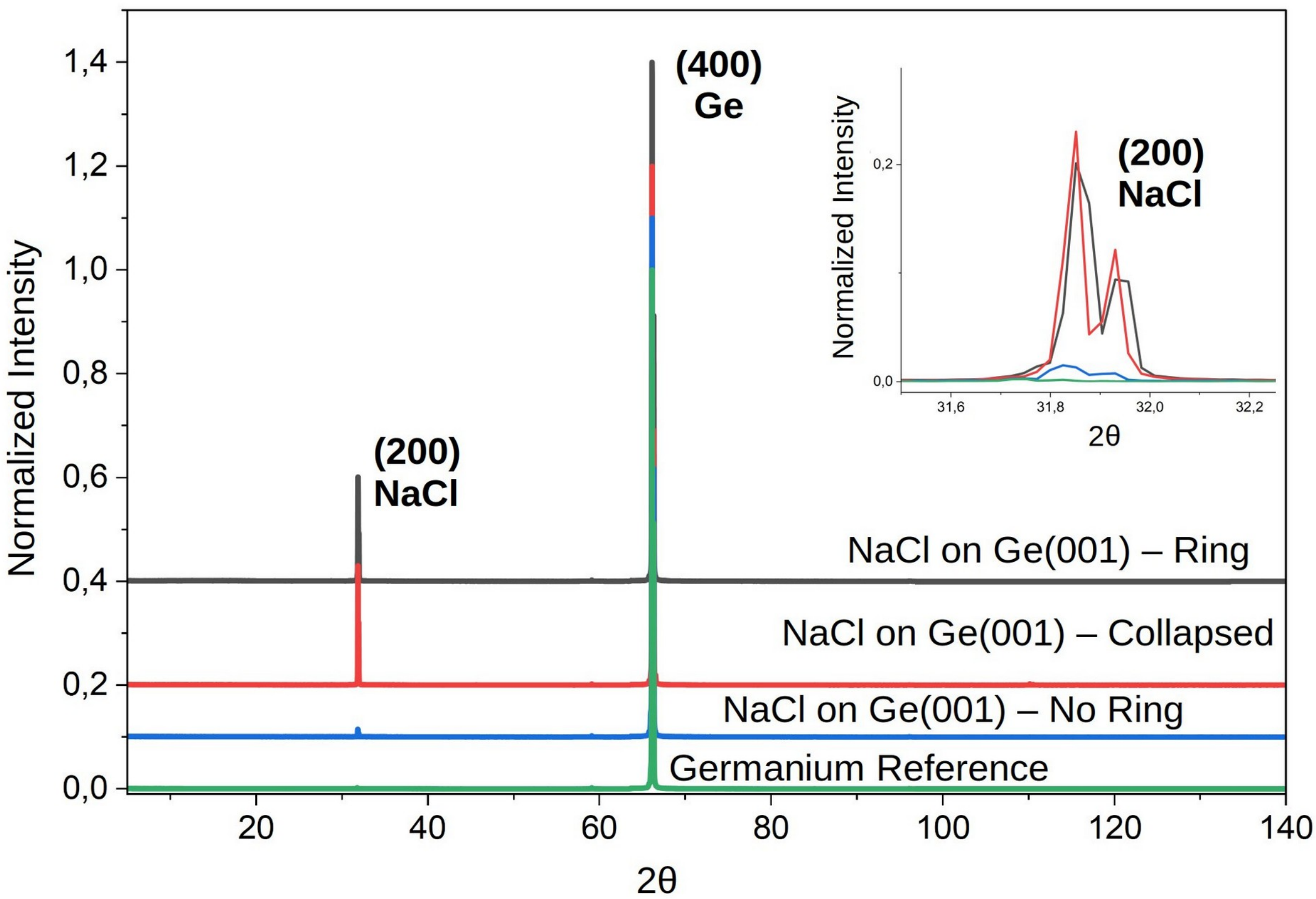

*Figure S3: XRD (θ–2θ) measurements of NaCl grown on a Ge(001) substrate. The diffraction patterns correspond to three distinct growth regimes: with coffee ring formation, with collapsed coffee ring, and without coffee ring. A reference pattern for crystalline germanium is also included. In all cases, in addition to the Ge(400) peak from the substrate, only the NaCl(200) peak is observed. This consistent diffraction signature unambiguously confirms the epitaxial relationship NaCl(001)//Ge(001) across all growth regimes.*

**Quantitative Thin Film Interference analysis**

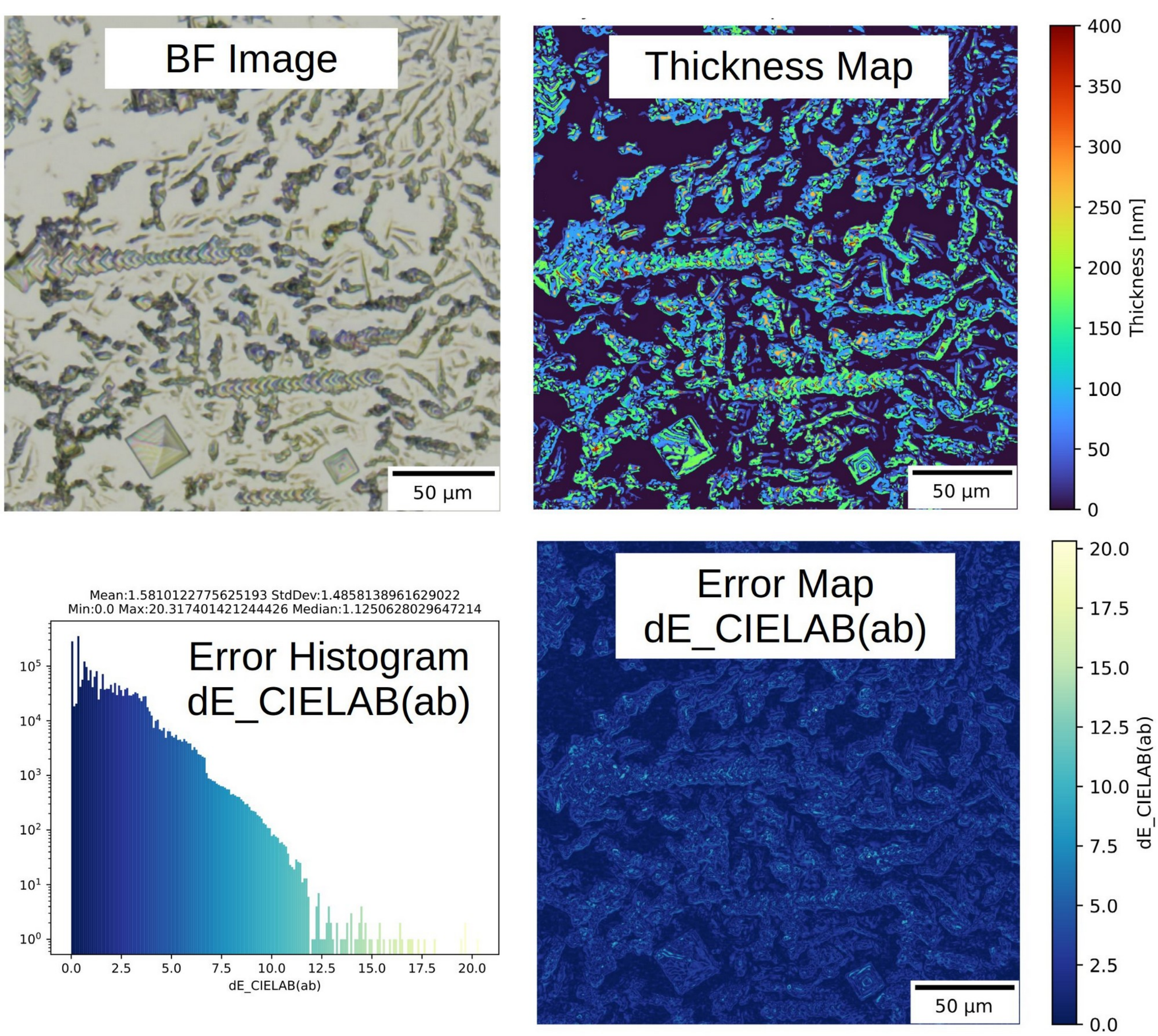

*Figure S4: Results of the Quantitative Thin Film Interference analysis on BF Image. Thickness map together with error map (dE_CIELAB(ab)) together with its histogram. It is worth to notice that average dE after fitting the model to the BF image is only 1.58 which is very good results.*

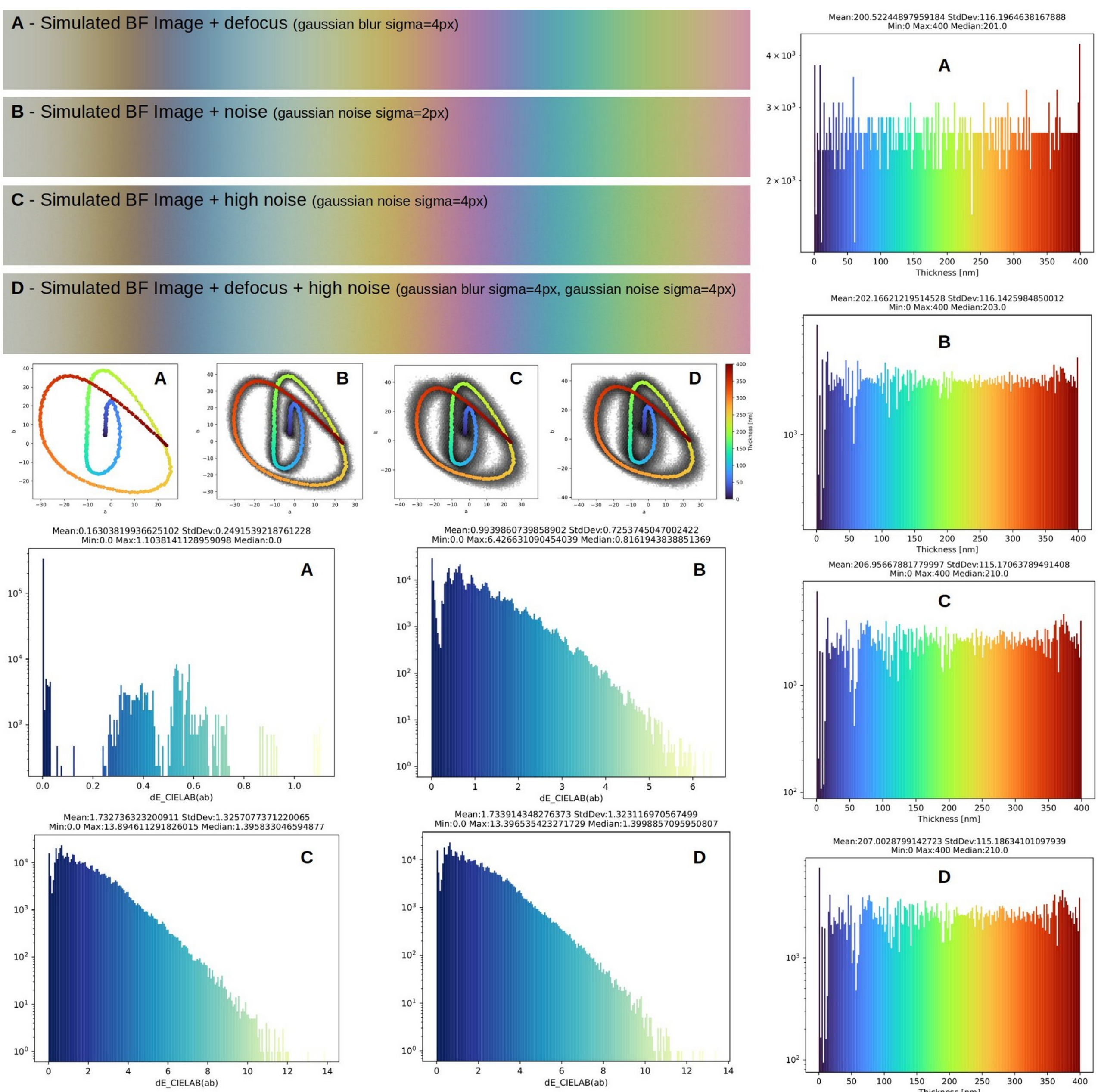

*Figure S5: Effect of the defocus and noise on the thickness determination by Quantitative Thin Film Interference analysis. Simulated BF images with homogeneous thickness distribution from 0 to 400nm corrupted by defocus A , noise B, high noise C and combined defocus and high noise D, as in real experiments. Corresponding CIELAB a-b space showing fitted model (color) and data (gray) together with dE_CIELAB(ab) error histograms and obtained thickness histograms from the fit of the model to the image. From the thickness histograms, it is seen that the most effected thickness range is from 0 to 25nm. It is also seen that for the worst case i.e. combined defocus with high noise the average thickness is 207.0nm (for ideal case one would expect 200nm) so the corresponding error related to defocus and high noise for the thickness determination via Quantitative Thin Film Analysis is estimated to be 3.5%.*

| **BF Image** | **Effective Estimated Noise Level on germanium substrate (std.dev. of intensity)** |
|---|---|
| Simulated BF + defocus A | 0.497 |
| Simulated BF + noise B | 1.238 |
| Simulated BF + high noise C | 2.312 |
| Simulated BF + defocus + high noise D | 2.433 |
| Experimental Raw BF Image | 0.996 |
| Experimental BF Image Processed<br>Median radius 4px as for the Thickness Analysis | 0.571 |

*Table S1: Comparison of effective estimated noise level on germanium substrate (std.dev. of intensity) for simulated and experimental image. It is seen that the experimental BF image which is used for the thickness analysis has a very low noise level (0.571) consistent with simulated BF image with defocus (0.497).*

## SEM Imaging and EBSD

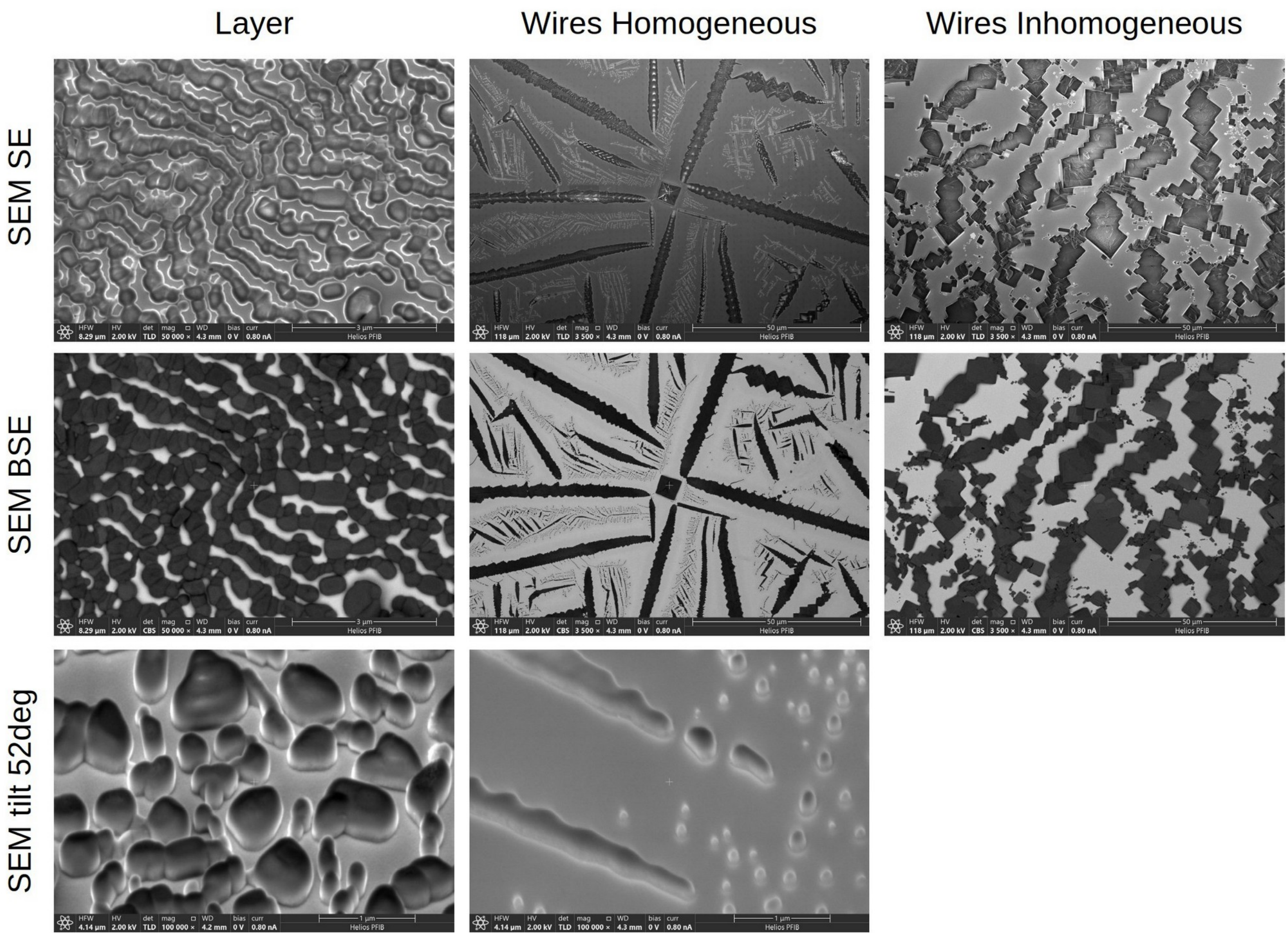

*Figure S6: Results of the SEM imaging of NaCl on Germanium (001) surface of three samples the homogeneous layer, the homogeneous wires and the inhomogeneous wires. The SEM SE shows the topographic contrast while SEM BSE shows compositional contrast (atomic number Z contrast). The thickness as determined from the SEM tilted by 52degree images agrees with the results from the Quantitative Thin Film Interference analysis.*

# Layer

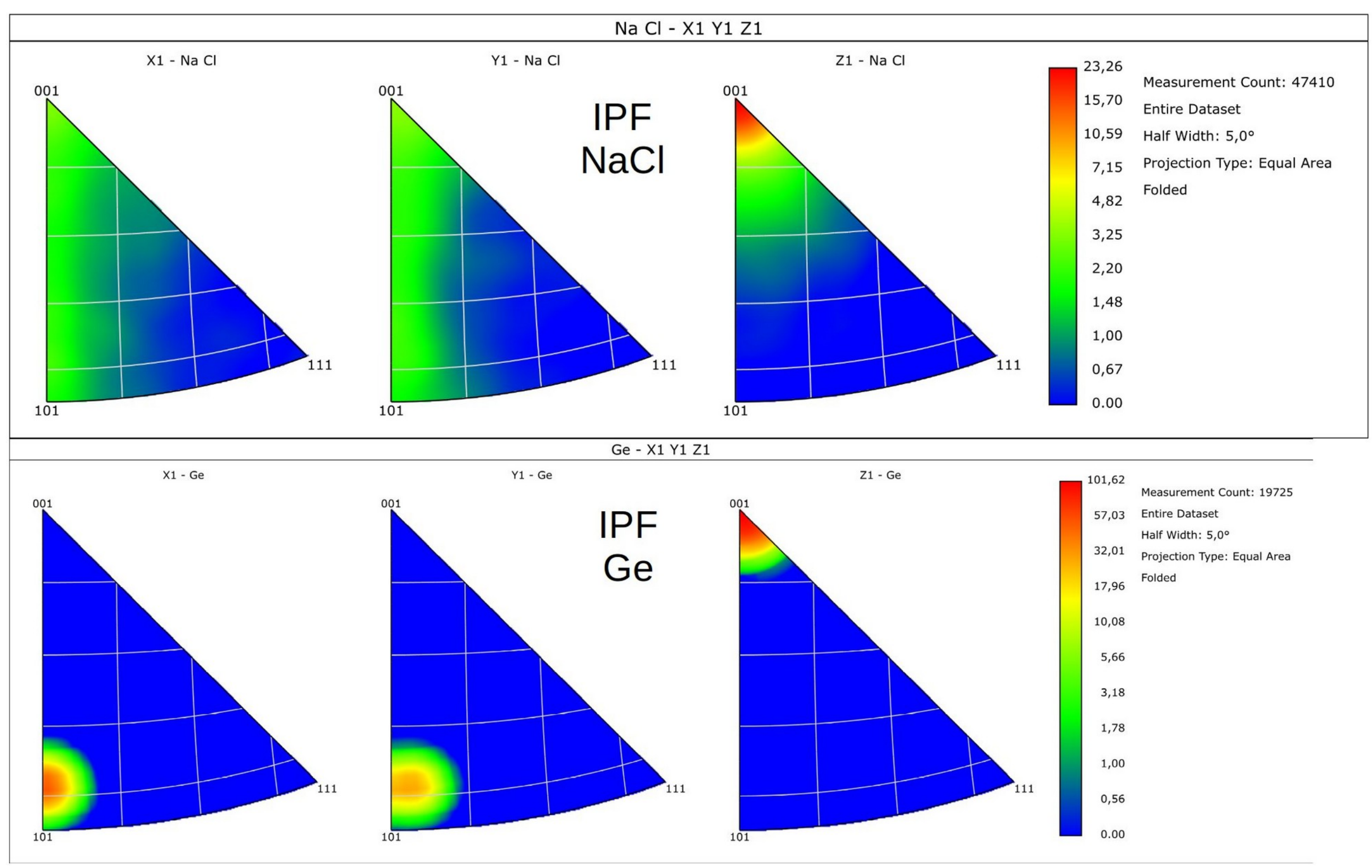

*Figure S7: NaCl on Germanium (001) surface. SEM EBSD Inverse Pole Figures (IPF) of Layer sample.*

# Homogeneous Wires

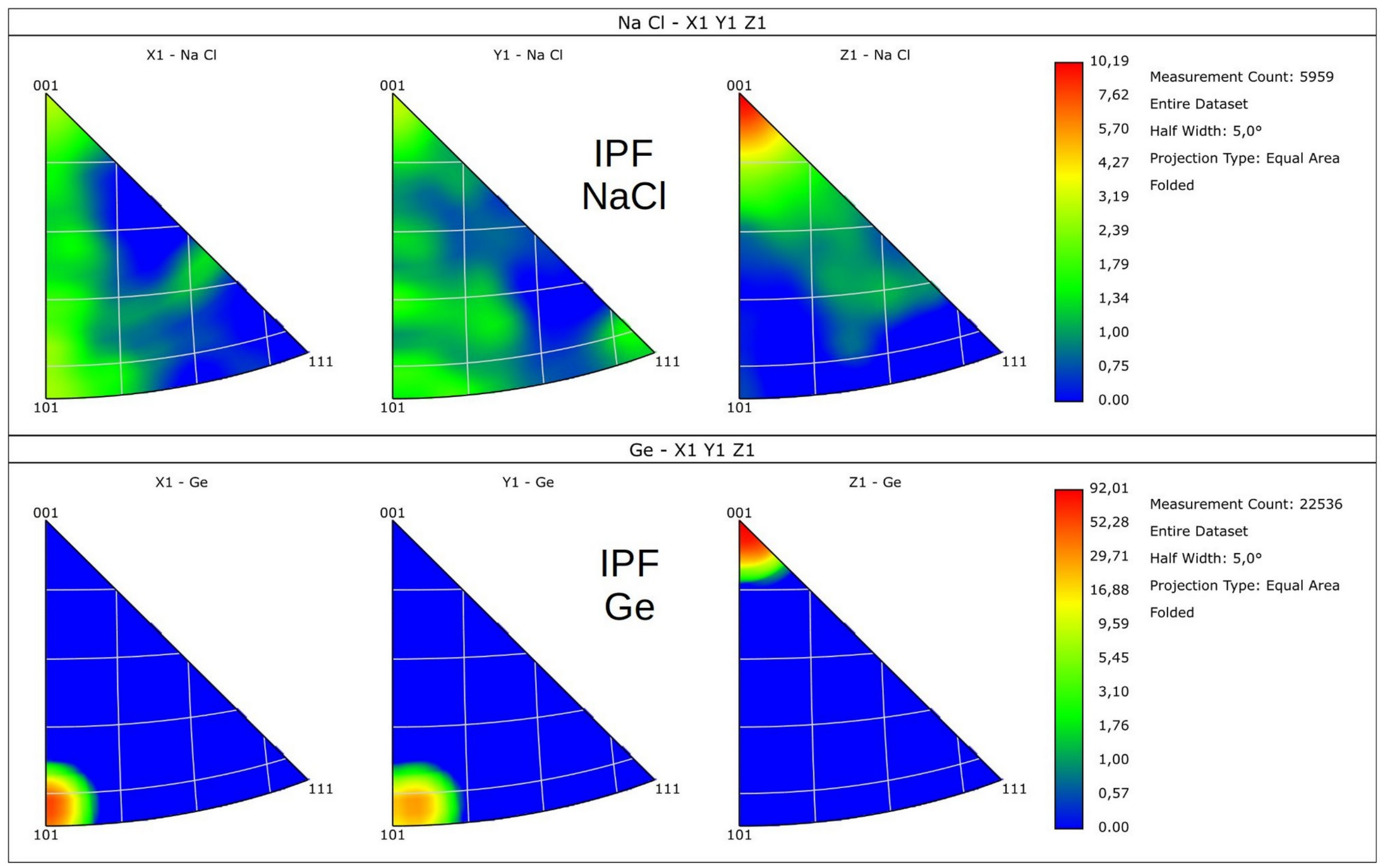

*Figure S8: NaCl on Germanium (001) surface. SEM EBSD Inverse Pole Figures (IPF) of homogeneous wires sample.*

# Inhomogeneous Wires

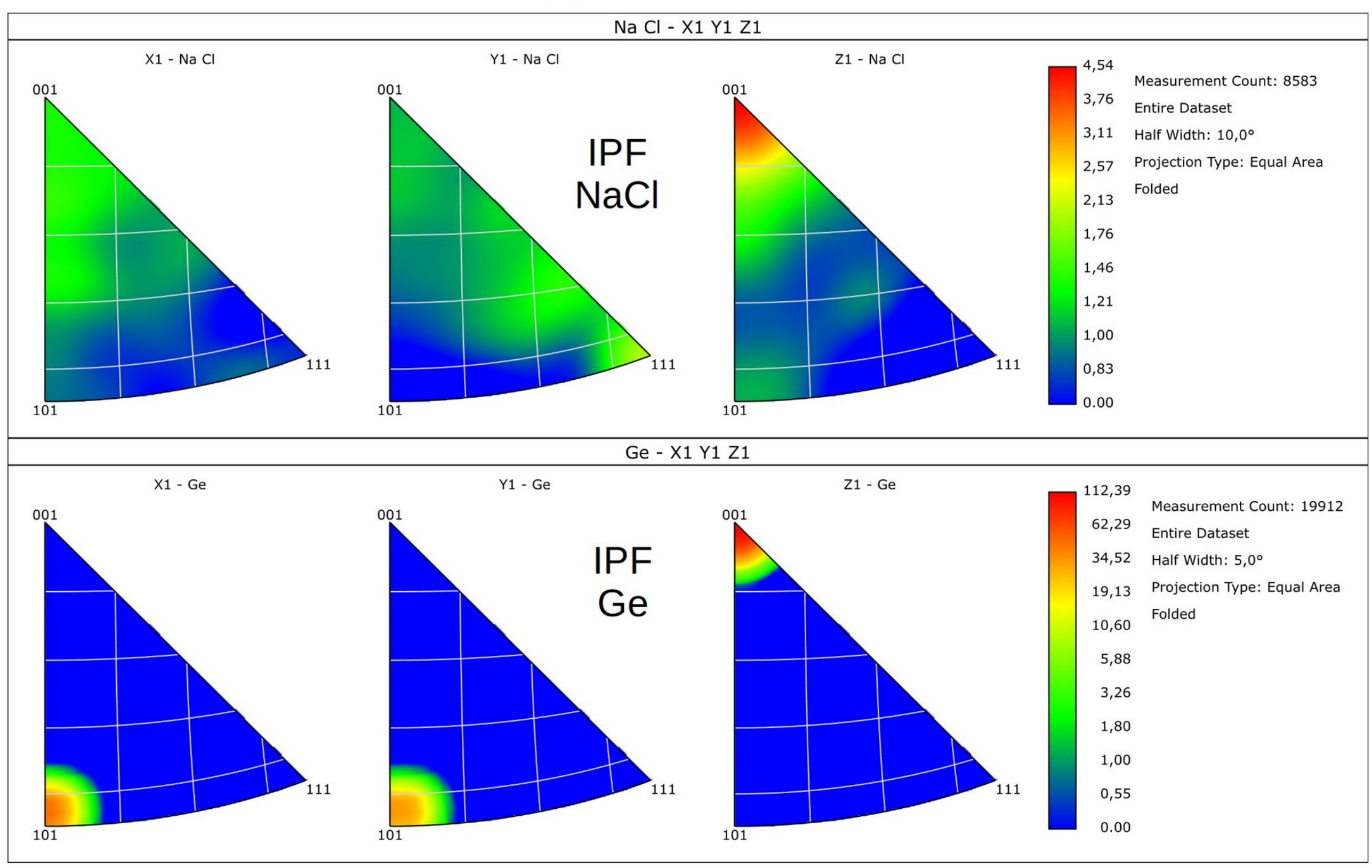

*Figure S9: NaCl on Germanium (001) surface. SEM EBSD Inverse Pole Figures (IPF) of inhomogeneous wires sample.*

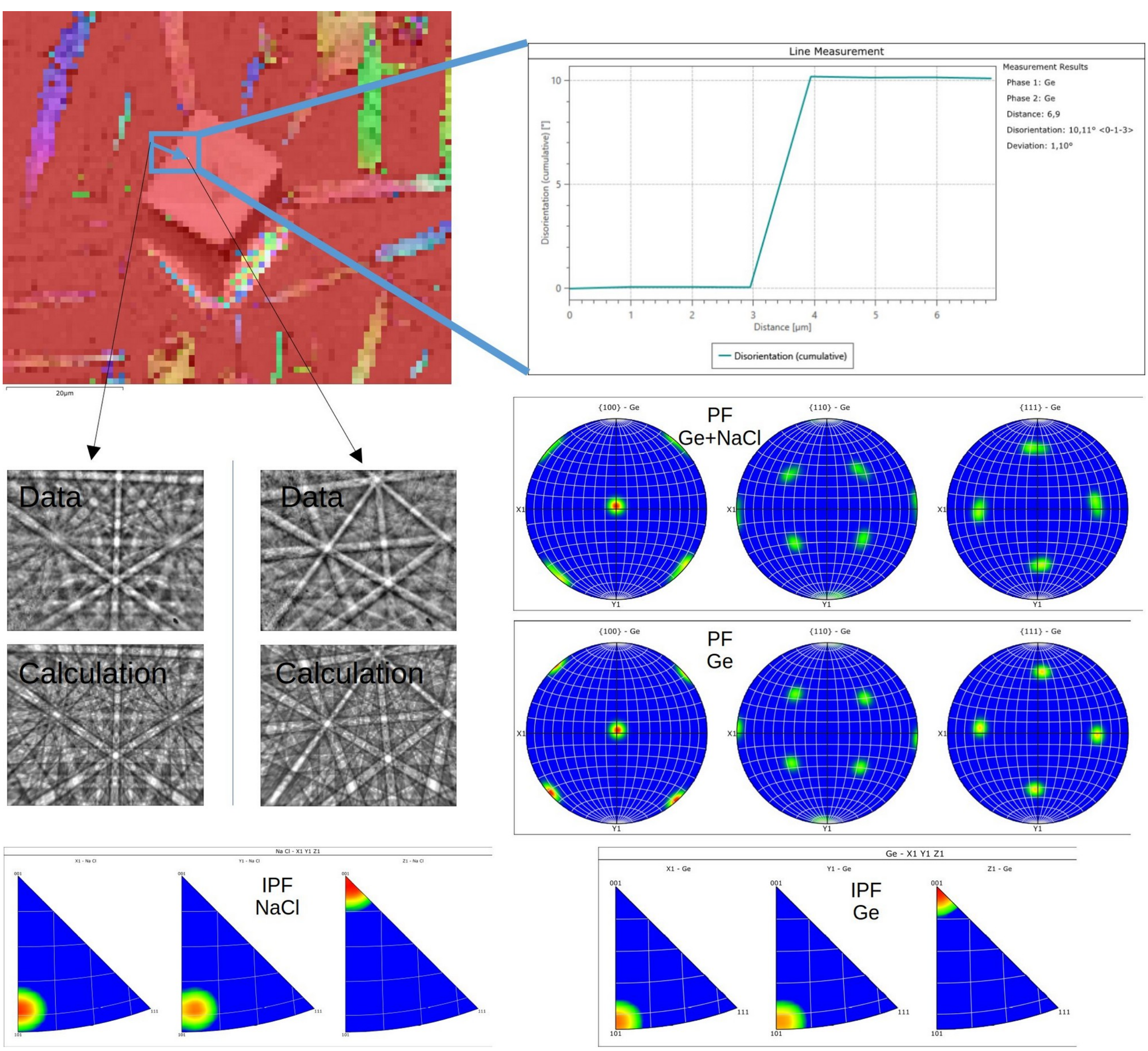

*Figure S10: Results of SEM EBSD (detail view – zoom). IPF Map of NaCl crystallite on germanium 001 surface together with EBSD patterns and misorientation plot. The Arrow on IPF map indicates the direction of the misorientation calculation. Its is seen that the NaCl crystallite is rotated "in plane" by 10.11 degree in respect the germanium substrate direction. This effect is also seen on the corresponding Pole Figures (PF) and Inverse Pole Figures (IPF).*

**Layer NaCl on Germanium (001) Surface**
**Bright Field and Dark Field Microscopy (Zoom – high magnification)**

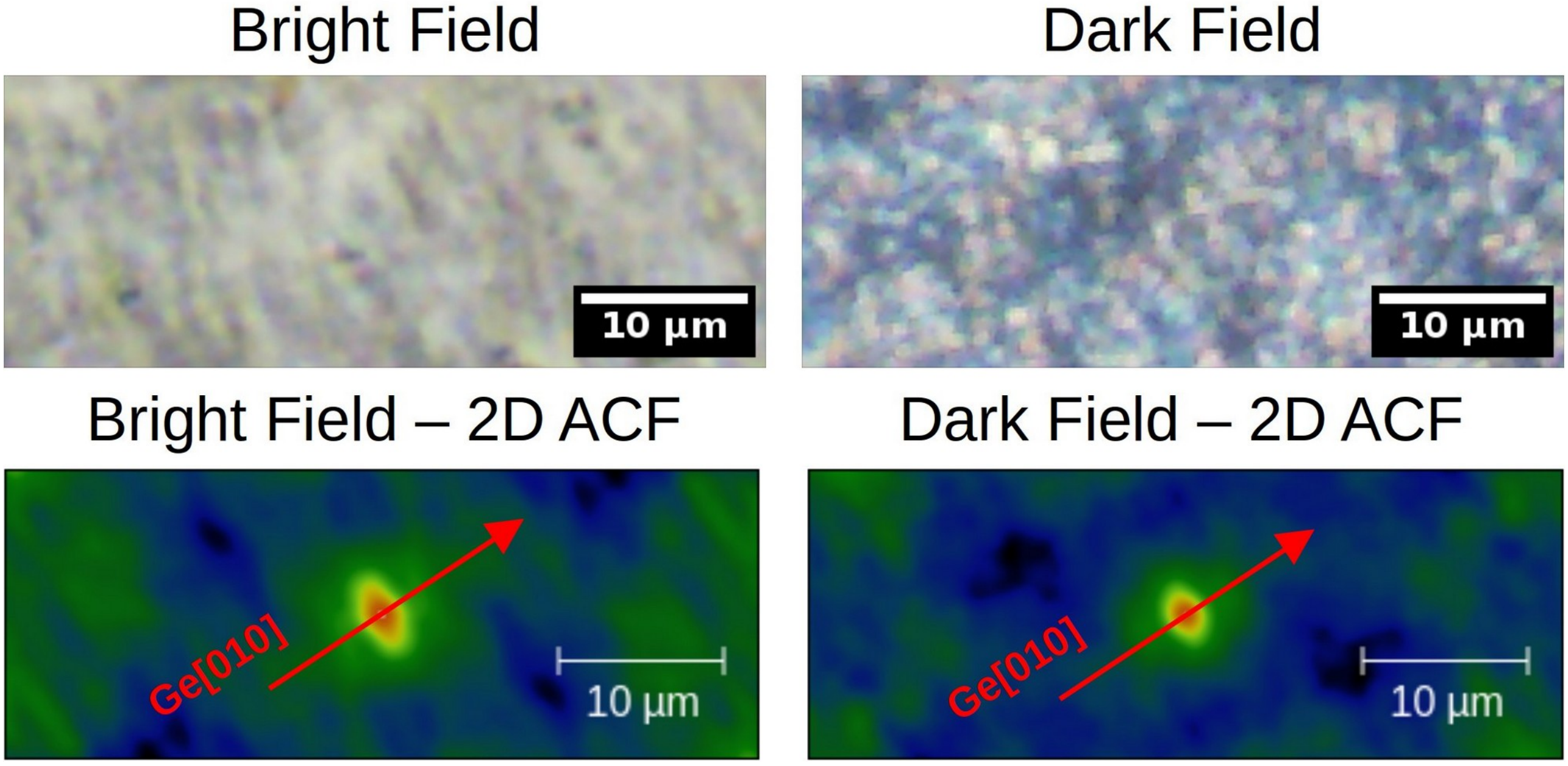

*Figure S11: Layer NaCl on Germanium (001) surface. Bright Field (BF) and Dark Field (DF) Microscopy Images (Zoom – high magnification) together with the corresponding 2D Autocorrelation Function (ACF). On the BF and DF images, particularly on 2D ACF, the asymmetry and the characteristic modulations along Ge[010] direction are visible, similarly as for the SEM imaging. Confirming the formation of 1D photonic crystal structure. Additionally in DF microscopy small grains like features are visible indicating a rough surface, similarly as for the SEM imaging.*

**NaCl Thickness Quantization**

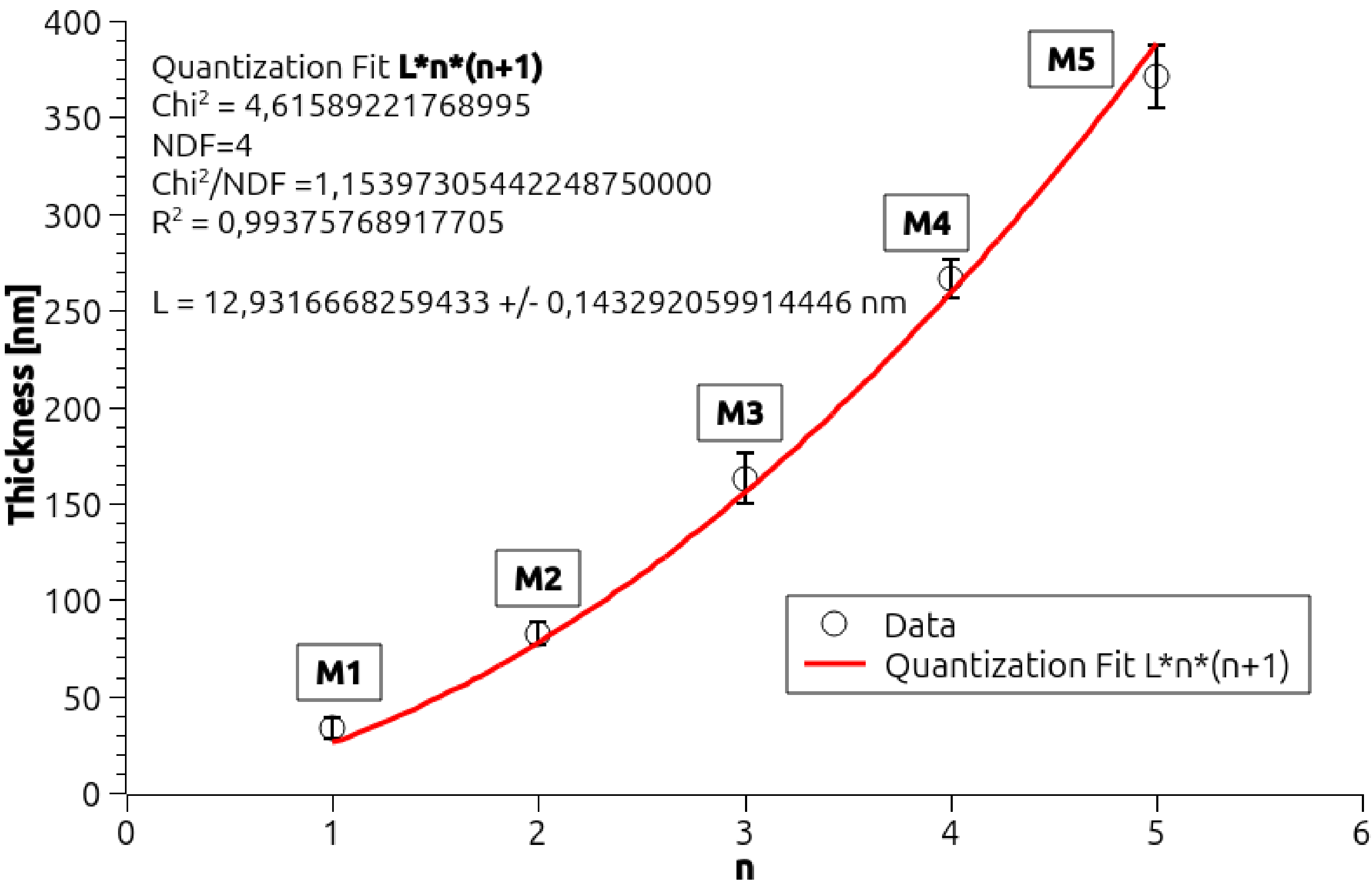

*Figure S12: Measured NaCl thin film thickness as a function on peak number n, together with quantization model fit L*n*(n+1), where L=12.93±0.14 nm is the effective relaxation length. The width of the thickness peaks was taken as measured thickness error. It is seen that the proposed model describes the experimental data very well $Chi^2/NDF=1.15$ and $R^2=0.9937$.*